\documentclass[12pt]{article}
\usepackage[left=2.5cm,right=2.5cm,top=4.45cm,bottom=4.45cm]{geometry}
\usepackage{ragged2e}
\usepackage[T1]{fontenc}
\usepackage[latin9]{inputenc}
\usepackage{babel}
\usepackage{floatrow}
\usepackage{float}
\usepackage{bm}
\usepackage[unicode=true]{hyperref}
\usepackage{amsmath}
\usepackage{amsthm}
\newtheorem{lemma}{Lemma}
\newtheorem{corollary}{Corollary}
\newtheorem{proposition}{Proposition}

\usepackage{amssymb}
\usepackage{graphicx}
\usepackage{xargs}

\usepackage{ifpdf}
\ifpdf
\else
\fi

\DeclareMathOperator{\ity}{iType}
\DeclareMathOperator{\depth}{depth}
\DeclareMathOperator{\dom}{dom}
\DeclareMathOperator{\iun}{iUniv}
\DeclareMathOperator{\isot}{isoType}
\DeclareMathOperator{\spec}{spec}
\DeclareMathOperator{\sg}{Sg}
\DeclareMathOperator{\rty}{relType}
\DeclareMathOperator{\pat}{pattern}

\DeclareMathOperator{\ar}{ar}

\DeclareMathOperator{\unk}{\mathtt{Null}}%

\DeclareMathOperator{\type}{\texttt{Type}}

\DeclareMathOperator{\true}{\textbf{True}}
\DeclareMathOperator{\false}{\textbf{False}}%
\DeclareMathOperator{\breakk}{\textbf{break}}%
\usepackage{changebar}
\providecommand{\lyxadded}[3]{}

\renewcommand{\lyxadded}[3]{
  {\protect\cbstart\color{lyxadded}{}#3\protect\cbend}
}

\newtheorem{thm}{Theorem}%[section]

\usepackage{tikz}
\usetikzlibrary{shapes}
\usetikzlibrary{plotmarks}
\usetikzlibrary{arrows}

\usetikzlibrary{shapes.geometric}

\usepackage[noend]{algorithm}
\usepackage[noend]{algpseudocode}
\newcommand\doubleplus{+\kern-1.3ex+\kern0.8ex}
\usepackage{xpatch}

\algrenewcommand\alglinenumber[1]{{\sffamily\footnotesize#1}}


\begin{document}

\global\long\def\itya#1{\ity(\bar{#1})}%

\newcommand{\dobleplus}{%
  \mathbin{{+}\mspace{-8mu}{+}}%
}

\global\long\def\iuna#1{\iun(\bar{#1})}%

\global\long\def\isota#1{\isot(\bar{#1})}%

\title{Two algorithms to decide Quantifier-free Definability in Finite Algebraic Structures}

\author{Miguel Campercholi\thanks{camper@famaf.unc.edu.ar} \and Mauricio Tellechea\thanks{mauriciotellechea@gmail.com} \and Pablo Ventura\thanks{pablogventura@gmail.com}}

\date{Universidad Nacional de C\'ordoba, C\'ordoba, Argentina}
%\history{(Received xx xxx xxx; revised xx xxx xxx; accepted xx xxx xxx)}

\maketitle

\justify

\begin{abstract}
This work deals with the definability problem by quantifier-free first-order formulas over a finite algebraic structure. We show the problem to be coNP-complete
and present two decision algorithms based on a semantical characterization of definable relations as those preserved by isomorphisms of substructures, the second one also providing a formula in the positive case.
Our approach also includes the design of an algorithm that computes
the isomorphism type of a tuple in a finite algebraic structure. Proofs of soundness
and completeness of the algorithms are presented, as well as empirical
tests assessing their performances.
\end{abstract}

\section{Introduction\label{sec:introduccion}}

Given a logic $\mathcal{L}$, a model-theoretic semantics for $\mathcal{L}$
is a function that assigns to each formula $\varphi$ of $\mathcal{L}$
and each structure $\mathbf{A}$ for $\mathcal{L}$ a set of tuples
from the domain of $\mathbf{A}$. This set of tuples is called the
\emph{extension }of $\varphi$ in $\mathbf{A}$  and we usually think
of these tuples as the interpretations that make $\varphi$ true in
$\mathbf{A}$, though it also makes sense to think of them as the
ones singled out, or named, by the properties expressed in $\varphi$.
Various fundamental computational problems arise from this setting,
of which surely the most prominent one is the \emph{satisfiability}
problem for $\mathcal{L}$, that for a given formula $\varphi$ from
$\mathcal{L}$ asks whether there is model $\mathbf{A}$ such that
the extension of $\varphi$ in $\mathbf{A}$ is non-empty. The \emph{model-checking}
problem for $\mathcal{L}$, which consists in computing, given a model
$\mathbf{A}$ and a formula $\varphi$, the extension of $\varphi$
in $\mathbf{A}$, has also played an important role in computational
logic. A third fundamental problem, the one that concerns us in this
article, is the \emph{definability }problem for $\mathcal{L}$, which
given a finite model $\mathbf{A}$ and a set $R$ of tuples from $\mathbf{A}$
asks whether there is a formula $\varphi$ such that the extension
of $\varphi$ in $\mathbf{A}$ agrees with $R$. This problem has
been studied for several logics, due to its fundamental nature and
its applications, for example to the Theory of Databases \cite{QueryByOutput}
and in the Generation of Referral Expressions \cite{krah:comp12,arec:refe08,arec:usin11}.
The computational complexity of the definability problem has also
been investigated for various logics. The article \cite{Kavvadias1998}
investigates the definability problem for classical propositional
logic under the name of \emph{inverse satisfiability}, proving it
is complete for coNP in the general case. They also characterize for
which special syntactical cases the problem lies in P. It is proved
in \cite{Aren:exac16} that the definability problem is GI-complete
(under Turing reductions) for first-order logic. For the primitive
positive fragment of first-order logic the definability problem is
coNEXPTIME-complete \cite{will:test10}, and for basic modal logic
\cite{arec:refe08} and some of its fragments \cite{arec:usin11}
it is in P. In the article \cite{complejidad}, it is shown that for
the quantifier-free fragment of first-order logic with a purely relational
vocabulary the definability problem is coNP-complete, and in the same
article a parameterized version of the problem is proved to be complete
for the complexity class W{[}1{]}.

To illustrate the concept of definability let us look at an example.
Let $\mathbf{D}=\langle\{\bot,\top,u,u'\},\wedge,\vee\rangle$ be the lattice
depicted below, and let $R=\{(a,b)\in D^{2}:a\leq b\}$.
\begin{center}
\begin{figure}[h]
\label{fig rombo}
\begin{centering}
\centering
\captionsetup{justification=centering}
\begin{tikzpicture}[line cap=round,line join=round,>=triangle 45,x=1.0cm,y=1.0cm]
    \clip(-3.14,-0.61) rectangle (3.01,2.56);
    \draw (-1,1)-- (0,2);
    \draw (0,2)-- (1,1);
    \draw (1,1)-- (0,0);
    \draw (0,0)-- (-1,1);
    \begin{scriptsize}
        \fill [color=black] (0,0) circle (1.5pt);
        \draw[color=black] (0.0,-0.25) node {$\bot$};
        \fill [color=black] (1,1) circle (1.5pt);
        \draw[color=black] (1.25,1.01) node {$u'$};
        \fill [color=black] (-1,1) circle (1.5pt);
        \draw[color=black] (-1.2,1.01) node {$u$};
        \fill [color=black] (0,2) circle (1.5pt);
        \draw[color=black] (0.01,2.2) node {$\top$};
    \end{scriptsize}
\end{tikzpicture}

\caption{The lattice $\mathbf{D}$}

\par\end{centering}

\end{figure}
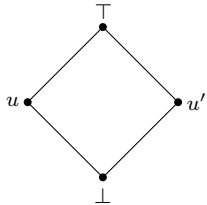
\par\end{center}

We want to know if $R$ is quantifier-free definable in $\mathbf{D}$.
That is, if there exists a quantifier-free formula $\varphi(x,y)$, in the
vocabulary of $\mathbf{D}$, such that $R=\{(a,b)\in D^{2}:\mathbf{D}\vDash\varphi(a,b)\}$.
It is easy to see that the formula $\varphi(x,y):=(x\vee y)=y$ works.
Next, consider the relation $R'=\{(\bot,u),(\bot,u'),(\bot,\top)\}$; is it quantifier-free definable
in $\mathbf{D}$? The answer is no, and here is why. Note that the
map $\gamma:\{\bot,\top\}\rightarrow\{u,\top\}$ is an isomorphism between
sublattices of $\mathbf{D}$, and quantifier free formulas are \emph{preserved}
by such maps. That is, if $\varphi(x,y)$ is quantifier free and $\mathbf{D}\vDash\varphi(a,b)$,
then $\mathbf{D}\vDash\varphi(\gamma a,\gamma b)$. Thus, for any
structure, every relation definable by a quantifier-free formula is
preserved by isomorphisms between substructures (we call these maps
subisomorphisms). Returning to our example, we can see that $(\bot,\top)\in R'$
but $(\gamma\bot,\gamma\top)\notin R'$, so $R'$ is not preseved by $\gamma$,
and thus it is not quantifier-free definable in $\mathbf{D}$. Interestingly,
the converse of this criterion holds for finite structures; i.e.,
if a $\mathbf{A}$ is a finite structure, then a relation $R\subseteq A^{k}$
is quantifier-free definable in $\mathbf{A}$ if and only if $R$
is preserved by all subisomorphisms of $\mathbf{A}$ (see Theorem \ref{thm:lema sem} below).

This characterization of quantifier-free definable relations is applied in \cite{relacional}
to obtain an algorithm that decides the definability problem
for the quantifier-free fragment of first-order logic with a purely
relational vocabulary (called the \emph{open-definability} problem
in \cite{relacional}). In the current note, which continues and complements the work in \cite{relacional}, we
develop two algorithms based on the same semantical characterization
but for algebraic vocabularies (i.e., without relation symbols).
It is worth noting that in fragments of first-order logic that allow
quantification, the relational and algebraic definability problems
collapse, since, at the expense of introducing quantifiers, every
formula can be effectively transformed into an equivalent one which
is unnested. However, in the quantifier-free case the problems are
not the same and require different algorithmic strategies. The main
difference lies in the fact that computing the isomorphisms
between substructures of a structure in the algebraic case, requires
computing the subuniverses of the structure. Our first approach in this article, the merging strategy to decide definability, is based on the
observation that computing the subuniverse generated by (the elements
of) a tuple $\bar{a}$ in a structure $\mathbf{A}$ is not far off
from computing the isomorphism type of $\bar{a}$ in a $\mathbf{A}$
(see Section \ref{sec:hit}). Our second approach, the splitting strategy, is also based on the same observation, but this time we take into account the target to allow us to prune part of the search space, and also generate a defining formula in the positive case.

In addition to presenting the above mentioned algorithms, we prove
that the computational complexity of the quantifier-free definability
problem for algebraic vocabularies is complete for coNP.

The paper is organized as follows. In the next section we fix notation
and provide the basic definitions. Section \ref{sec:complejidad} is devoted to the exposition of our complexity result. In Section \ref{sec:hit} we introduce the algorithm
that computes the isomorphism type of a tuple in an algebra. The merging strategy algorithm to decide quantifier-free-definability for algebraic structures is presented
in Section \ref{sec:megahit}, while the splitting strategy algorithm, which also allows to compute formulas, is presented in Section \ref{sec:posta}. In Section \ref{sec:testing} some empirical tests are performed, comparing the performance of our two algorithmic strategies. These tests also show that our algorithms perform better on models with a large amount of symmetries. Finally, Section \ref{sec:conclusiones} provides a summary of the results obtained as well as further research directions which arise from the developments we present here.

\section{Preliminaries\label{sec:preliminares}}

In this section we provide some basic definitions and fix notation.
We assume basic knowledge of first-order logic. For a detailed account
see, e.g., \cite{ebbi:math}. Given a first order vocabulary $\tau$
and $k\in\omega$ let $\mathcal{T}_{k}$ denote the set of $\tau$-terms
of depth $k$ with variables from $\textsf{VAR}:=\{x_{0},x_{2},\ldots,x_{n},\ldots\}$. We also use $\mathcal{T}_k (v_{0},\ldots,v_{k})$ when the variables are in $\{ v_{0},\ldots,v_{k} \}$. We write $\mathcal{T}:=\bigcup_{k\in\omega}\mathcal{T}_{k}$ for the set of all $\tau$-terms over $\textsf{VAR}$. For a function or relation
symbol $s$ we write $\ar(s)$ denote the arity of $s$. In this
note we are concerned with \emph{algebraic} vocabularies, that is,
without relation symbols. In this context, \emph{atomic formulas}
are of the form $t=t'$ where $t$ and $t'$ are terms. A \emph{quantifier-free} formula is a boolean combination of atomic
formulas. We write $\varphi(v_{0},\ldots,v_{k})$ for a formula $\varphi$
whose variables are all included in $\{v_{0},\ldots,v_{k}\}$, and a similar notation is used for a term $t(v_{0},\ldots,v_{k})$. An
\emph{algebra} is a structure of an algebraic vocabulary; algebras
are denoted by boldface letters (e.g., $\mathbf{A},\mathbf{B},\mathbf{C},$...)
and their universes by the corresponding non-bold letter. Given an
algebra $\mathbf{A}$, a formula $\varphi(v_{0},\ldots,v_{k})$, and
a sequence of elements $\bar{a}=(a_{0},\ldots,a_{k-1})\in A^{k}$
we write $\mathbf{A}\models\varphi(\bar{a})$ if $\varphi$ is true in
$\mathbf{A}$ under an assignment that maps $v_{i}$ to $a_{i}$.
We use $[\varphi(\bar{v})]^\mathbf{A}$ (or simply $[\varphi]^\mathbf{A}$ when no confusion may arise) to denote the extension of $\varphi$ in $\mathbf{A}$. We also say that a subset $R\subseteq A^{k}$ is \emph{quantifier-free definable}
in \textbf{A} (\emph{qf-definable} for short) provided there is a quantifier-free first-order formula $\varphi(x_{0},\dots,x_{k-1})$
in the vocabulary of $\mathbf{A}$ such that 
\[
R=[\varphi]^\mathbf{A}=\{\bar{a}\in A^{k}:\bm{A}\models\varphi(\bar{a})\}.
\]
In this article we study the following computational decision problem:\medskip{}

\noindent $\mathrm{QfDefAlg}$

\noindent \emph{Input}: A finite algebra $\mathbf{A}$ and a target
relation $R\subseteq A^{k}$.

\noindent \emph{Question}: Is $R$ qf-definable in $\mathbf{A}$?\medskip{}

Let $\tau$ be an algebraic vocabulary and $\mathbf{A}$ be an algebra. A subset $S\subseteq A$ is a \emph{subuniverse} of $\mathbf{A}$ if
$S$ is nonempty and it is closed under the fundamental operations of $\mathbf{A}$.
For a tuple $\bar{a}=(a_{0},\ldots,a_{k-1})\in A^{k}$
we write $\sg^\mathbf{A}(\bar{a})$ (or just $\sg(\bar{a})$ when the context is clear) to denote the subuniverse of $\mathbf{A}$ generated by $\{a_{0},\ldots,a_{k-1}\}$.
Let $\gamma:\dom\gamma\subseteq A\rightarrow A$ be a function. We say that $\gamma$ \emph{preserves} $R\subseteq A^k$ if for all $(a_{0},\dots,a_{k-1}) \in R\cap(\dom\gamma)^{k}$
we have $(\gamma a_{0},\dots,\gamma a_{k-1}) \in R$.
The function $\gamma$ is a \emph{subisomorphism} of $\mathbf{A}$
provided that $\gamma$ is injective, $\dom\gamma$ is a subuniverse of $\mathbf{A}$,
and $\gamma$ preserves the graph of $f^{\mathbf{A}}$ for each $f\in\tau$.
(Note that a subisomorphism of $\mathbf{A}$ is precisely an isomorphism
between two substructures of $\mathbf{A}$.)
The following result characterizing qf-definability in terms of subisomorphisms plays a central role in this article.

\begin{thm}
\cite[Thm 3.1]{camp:lemas_semanticos}
\label{thm:lema sem} Let $\mathbf{A}$ be a finite model of an arbitrary
first order signature and let $R\subseteq A^{k}$. The following are
equivalent:

\begin{itemize}
\item $R$ is definable in $\mathbf{A}$ by a quantifier-free first-order
formula.
\item $R$ is preserved by all isomorphisms between $k$-generated substructures
of $\mathbf{A}$.
\end{itemize}
\end{thm}

\section{$\mathrm{QfDefAlg}$ is coNP-complete \label{sec:complejidad}}

In what follows a \emph{graph} is a model $\mathbf{G}$ of the language
with a single binary relation $E$, such that $E^{\mathbf{G}}$ is
symmetric and irreflexive. In \cite{complejidad} it is proved that
the following problem is complete for $\mathrm{coNP}$.\medskip{}

\noindent $\mathrm{QfDef[Graphs]}$

\noindent \emph{Input}: A finite graph $\mathbf{G}$ and a target
relation $R\subseteq G^{k}$.

\noindent \emph{Question}: Is $R$ qf-definable in $\mathbf{G}$?

\medskip{}
A \emph{subisomorphism} of a graph $\mathbf{G}=(G,E)$ is an injective function
$\gamma:\dom\gamma\subseteq G\rightarrow G$ such that $(a,b)\in E$ if and only if $(\gamma a,\gamma b)\in E$.

\begin{thm}

\label{ODA is coNP completo}$\mathrm{QfDefAlg}$ is $\mathrm{coNP}$-complete.

\end{thm}

\begin{proof}
Given a finite graph $\mathbf{G}=(G,E)$ let $\hat{0}$ and $\hat{1}$
denote the first two positive integers not in $G$. We define the
algebra $\mathbf{G}_{*}:=(G\cup\{\hat{0},\hat{1}\},f)$ where $f$
is the binary operation 

\[
f(a,b)=\begin{cases}
\hat{1} & \text{if }(a,b)\in E\text{ or }a=b\in\{\hat{0},\hat{1}\}\\
\hat{0} & \text{otherwise.}
\end{cases}
\]
It is straightforward to check that:
\begin{enumerate}
\item If $\gamma$ is a subisomorphism of $\mathbf{G}$, then the extension $\gamma_{*}$ of
$\gamma$ given by $\gamma_{*}(\hat{0})=\hat{0}$ and $\gamma_{*}(\hat{1})=\hat{1}$
is a subisomorphism of $\mathbf{G}_{*}$.
\item If $\delta$ is a subisomorphism of $\mathbf{G}_{*}$, then $\delta[G]\subseteq G$ and
$\delta|_{G}$ is a subisomorphism of $\mathbf{G}$.
\end{enumerate}
We prove next that $\left\langle \mathbf{G},R\right\rangle \longmapsto\left\langle \mathbf{G}_{*},R\right\rangle $
is a polynomial time (Karp) reduction from $\mathrm{QfDef[Graphs]}$
to $\mathrm{QfDefAlg}$. Clearly $\mathbf{G}_{*}$ can be computed
from $\mathbf{G}$ in polynomial time, so it remains to show that
$R$ is qf-definable in $\mathbf{G}$ if and only if $R$ is qf-definable
in $\mathbf{G}_{*}$. Fix a finite graph $\mathbf{G}$ and $R\subseteq G^{k}$.
Suppose $R$ is not qf-definable in $\mathbf{G}$. Then, by Theorem
\ref{thm:lema sem}, there is $\gamma$, a subisomorphism of $\mathbf{G}$, that does
not preserve $R$. Now (1) says that $\gamma_{*}$ is a subisomorphism
of $\mathbf{G}_{*}$, and since it does not preserve $R$, it follows
that $R$ is not qf-definable in $\mathbf{G}_{*}$. For the remaining
direction, suppose $R$ is not qf-definable in $\mathbf{G}_{*}$.
Once again Theorem \ref{thm:lema sem} produces a subisomorphism $\delta$
of $\mathbf{G}_{*}$ that does not preserve $R$. It follows from
(2), that $\delta|_{G}$ is a subisomorphism of $\mathbf{G}$ that
does not preserve $R$; hence $R$ is not qf-definable in $\mathbf{G}$.

Showing that $\mathrm{QfDefAlg}$ is in $\mathrm{coNP}$ is a straightforward
application of Theorem \ref{thm:lema sem}. In fact, each negative
instance $\left\langle \mathbf{A},R\right\rangle $ of $\mathrm{QfDefAlg}$
is witnessed by a bijection $\gamma$ between subsets of $A$ satisfying
conditions easily checked in poly-time w.r.t. the size of $\left\langle \mathbf{A},R\right\rangle $.

\end{proof}

It follows from the proof of Theorem \ref{ODA is coNP completo} that
the restriction of $\mathrm{QfDefAlg}$ to structures with a single
binary commutative operation is already complete for $\mathrm{coNP}$.

\section{Computing the isomorphism type of a tuple\label{sec:hit}}

In this section we present an algorithm to compute the isomorphism
type of a tuple $\bar{a}$ in a finite algebraic structure $\mathbf{A}$,
and prove it to be correct. We start with some needed definitions
and preliminary results.

Let $\mathbf{A}$ be a finite algebra, for each $k\in\omega$ we define
the equivalence relation $\approx_{k}$ over $A^{n}$ by $\bar{a}\approx_{k}\bar{b}$
if and only if for all terms $t,s\in\mathcal{T}_{k}(x_{0},\ldots,x_{n-1})$
we have that 
\[
t^{\mathbf{A}}(\bar{a})=s^{\mathbf{A}}(\bar{a})\Longleftrightarrow t^{\mathbf{A}}(\bar{b})=s^{\mathbf{A}}(\bar{b}).
\]
We say two tuples $\bar{a},\bar{b} \in A^n$ are \emph{isomorphic} (denoted by $\bar{a} \approx \bar{b}$) provided that $\bar{a}\approx_{k}\bar{b}$ for all $k\in\omega$. Notice $\approx$ is an equivalence relation over $A^{n}$.
For $\bar{a}\in A^{n}$ let
\[
\mathrm{K}_{\bar{a}}:=\min\{k\in\omega:\text{ for all \ensuremath{t\in\mathcal{T}_{k}}}(\bar{x})\text{ there is }\hat{t}\in\mathcal{T}_{k-1}(\bar{x})\text{ such that }t^{\mathbf{A}}(\bar{a})=\hat{t}^{\mathbf{A}}(\bar{a})\}.
\]
Note that, since $\mathbf{A}$ is finite, this minimum always exists.

\begin{lemma}

\label{lema mat}Let $\mathbf{A}$ be a finite algebra and $\bar{a},\bar{b}\in A^{n}$.
\begin{enumerate}
\item \label{repr comun}If $\bar{a}\approx_{\mathrm{K}_{\bar{a}}}\bar{b}$,
then for any term $t(\bar{x})$ there is a term $\hat{t}(\bar{x})\in\mathcal{T}_{l}$
with $l<\mathrm{K}_{\bar{a}}$ such that $t^{\mathbf{A}}(\bar{a})=\hat{t}^{\mathbf{A}}(\bar{a})$
and $t^{\mathbf{A}}(\bar{b})=\hat{t}^{\mathbf{A}}(\bar{b})$.
\item \label{isoK implica iso}If $\bar{a}\approx_{\mathrm{K}_{\bar{a}}}\bar{b}$,
then $\bar{a}\approx\bar{b}$.
\end{enumerate}
\end{lemma}

\begin{proof}(\ref{repr comun}) Observe first that from the definition
it follows that $\mathrm{K}_{\bar{a}}=\mathrm{K}_{\bar{b}}$; we write
$K$ for $\mathrm{K}_{\bar{a}}$. Fix a term $t(\bar{x})$. If $t(\bar{x})\in\mathcal{T}_{K-1}$
we can take $t=\hat{t}$, so suppose $t\in\mathcal{T}_{K+j}$ with
$j\geq0$. We show that $\hat{t}$ exists by induction on $j$. Suppose
$j=0$; by the definition of $\mathrm{K}_{\bar{a}}$ there is $\hat{t}\in\mathcal{T}_{K-1}$
such that $t^{\mathbf{A}}(\bar{a})=\hat{t}^{\mathbf{A}}(\bar{a})$,
and since $\bar{a}\approx_{\mathrm{K}}\bar{b}$ it follows that $t^{\mathbf{A}}(\bar{b})=\hat{t}^{\mathbf{A}}(\bar{b})$.
Assume next that $j>0$ and suppose $t=f(t_{0},...,t_{r-1})$, with
each $t_{i}\in\mathcal{T}_{K+j-1}$. By the inductive hypothesis,
there are terms $\hat{t}_{0},...,\hat{t}_{r-1}\in\mathcal{T}_{u}$ with
$u<K$, each satisfying $t_{i}^{\mathbf{A}}(\bar{a})=\hat{t_{i}}^{\mathbf{A}}(\bar{a})$
and $t_{i}^{\mathbf{A}}(\bar{b})=\hat{t_{i}}^{\mathbf{A}}(\bar{b})$.
That means $\tilde{t}:=f(\hat{t}_{0},...,\hat{t}_{r-1})\in\mathcal{T}_{u+1}$
satisfies $\tilde{t}^{\mathbf{A}}(\bar{a})=t^{\mathbf{A}}(\bar{a})$
and $\tilde{t}^{\mathbf{A}}(\bar{b})=t^{\mathbf{A}}(\bar{b})$. Since
$u+1\leq K$, by the case $j=0$ we have a term $\hat{t}\in\mathcal{T}_{l}$
with $l<K$ for which $\hat{t}^{\mathbf{A}}(\bar{a})=\widetilde{t}^{\mathbf{A}}(\bar{a})=t^{\mathbf{A}}(\bar{a})$
and $\hat{t}^{\mathbf{A}}(\bar{b})=\tilde{t}^{\mathbf{A}}(\bar{b})=t^{\mathbf{A}}(\bar{b})$.

(\ref{isoK implica iso}) Let $t,s\in\mathcal{T}$ such that $t^{\mathbf{A}}(\bar{a})=s^{\mathbf{A}}(\bar{a})$.
Take $\hat{t},\hat{s}$ the terms given by (\ref{repr comun}). Since
$\hat{t}^{\mathbf{A}}(\bar{a})=\hat{s}^{\mathbf{A}}(\bar{a})$, by
the hypothesis we have $\hat{t}^{\mathbf{A}}(\bar{b})=\hat{s}^{\mathbf{A}}(\bar{b})$,
and therefore $t^{\mathbf{A}}(\bar{b})=s^{\mathbf{A}}(\bar{b})$.

\end{proof}

Next we introduce, in the form of pseudocode, the algorithm that computes
the function isoType (see Algorithm \ref{alg:hit} below). This function takes 
as input a finite algebra $\mathbf{A}$ and a tuple $\bar{a}$ from $A$ and returns a 
representation of the isomorphism type of $\bar{a}$ together with the subuniverse of $\mathbf{A}$ 
generated by $\bar{a}$ listed in a canonical order. The isomorphism type is obtained by traversing 
the terms in a specified order and evaluating them on $\bar{a}$. As we shall see (Corollary \ref{correccion_hit}), 
it suffices to record which terms return the same value to capture the isomorphism type of $\bar{a}$.

Let us describe how Algorithm \ref{alg:hit} works. Fix an input algebra $\mathbf{A}$
with vocabulary $\tau$ (assume the function symbols in $\tau$ are
listed in a specific order). Given a tuple $\bar{a}\in A^{n}$, at
the start of the algorithm, variable $\mathtt{V}$ is set to $[a_{0},\ldots,a_{n-1}]$.
Then the while loop starts, and the elements added to $\mathtt{V}$
are obtained by applying fundamental operations of $\mathbf{A}$ to
earlier values in $\mathtt{V}$. Thus, all elements in $\mathtt{V}$
belong to $\sg(\bar{a})$. Variable $\mathtt{P}$ records the repetitions
in appearances of such elements as they are computed, while variable
$\mathtt{H}$ stores the indexes where each element of $\sg(\bar{a})$
appears for the first time in $\mathtt{V}$. At the end of each pass
through the while loop, variable $\mathtt{N}$ is assigned with the
indexes that were added to $\mathtt{H}$ on that pass. Since $A$
is finite, eventually no new elements are produced, and $\mathtt{N}$
is assigned with the empty list. This guarantees termination and the
fact that all elements of $\sg(\bar{a})$ are listed in $\mathtt{V}$
when the algorithm halts. Observe that the while loop is executed
exactly $\mathrm{K}_{\bar{a}}$ times on input $\bar{a}$. We remark that 
lines \ref{initV'} and \ref{compute term} - \ref{endvv'} in Algorithm \ref{alg:hit},
whose purpose is to initialize and update the variable $\mathtt{V'}$,
are not actually needed to compute the function isoType, but are included
to make our proofs easier to follow. 

To compute elements in $\mathtt{V}$ on each pass, for each arity
of some fundamental operation we construct a list $\mathtt{T}$ of
tuples of indexes of elements where the fundamental operations of
that arity will be applied. We require these tuples have at least
one element that has not appeared before to avoid unnecessary computations. 

\begin{algorithm}[H]
\caption{}
\label{alg:hit}

\begin{algorithmic}[1]

\Function{$\mathtt{IsoType}$}{$\mathbf{A}$, $\bar{a}$}

	\Comment{$\mathbf{A}$ is an algebra and $\bar{a}$ is a tuple from
$A$}

	\State{$\mathtt{V}=[a_{0},\dots,a_{n-1}]$}

	\State{$\mathtt{V'}=[``x_{0}",\dots,``x_{n-1}"]$} \Comment{this
is a list of variables (terms are represented as strings)}\label{initV'}

	\State{$\mathtt{P}=[B_{1},\dots,B_{k}]$ the partition of $\{0,\dots,n-1\}$
where $i,j$ are in the same block iff $a_{i}=a_{j}$}

	\State{$\mathtt{H}=\mathtt{sorted}([\min(B_{j}):B_{j}\in\mathtt{P}])$} \Comment{sorted
increasingly}

	\State{$\mathtt{N=H}$}

	\State{$\mathtt{arities=}[r_{1},...,r_{k}]$} \Comment{ where
$r_{1}<...<r_{k}$ are all the arities of operations in $\mathbf{A}$}

	\While{$\mathtt{N}\neq\emptyset$}

		\State{$\mathtt{H\_old=copy(H)}$}

		\For{$r\in\mathtt{arities}$}\label{fixr}

			\State{$\mathtt{T=sorted}([\bar{l}\in(\mathtt{H\_old})^{r}:l_{j}\in\mathtt{N}\text{ for some }j])$} \Comment{sorted
lexicographically}

			\For{$f\in\mathtt{op}(r)$} \Comment{$\mathtt{op}(r)$ the
list of ops symbols of arity $r$ in the order given by $\tau$}\label{fixf}

				\For{$\bar{l}\in\mathtt{T}$}\label{fixtup}

					\State{$\mathtt{value}=f^{\boldsymbol{A}}(\mathtt{V}[l_{0}],...,\mathtt{V}[l_{r-1}])$}\label{beginvv'}

					\State{append \texttt{value} to \texttt{V}}

					\State{$\mathtt{term}=f\dobleplus``("\dobleplus\mathtt{V'}[l_{0}]\dobleplus``,"\dobleplus...\dobleplus``,"\dobleplus\mathtt{V'}[l_{r-1}]\dobleplus``)"$}\label{compute term}

					\State{append \texttt{term} to \texttt{V$'$}}\label{endvv'}

					\Comment{update of the partition \texttt{P}}

					\If{there is $j\in\mathtt{H}$ such that $\mathtt{value}=\mathtt{V}[j]$} \Comment{\texttt{value}
was already in \texttt{V}}

						\State{add $|\mathtt{V}|-1$ to the block of \texttt{P} that
contains $j$}

					\Else \Comment{\texttt{value} is new}

						\State{add the block $\{|\mathtt{V}|-1\}$ to \texttt{P}}

						\State{add $|\mathtt{V}|-1$ to \texttt{H}}

					\EndIf

				\EndFor

			\EndFor

		\EndFor

		\State{$\mathtt{N}=[l:l\in\mathtt{H}\text{ and }l\notin\mathtt{H\_old}]$}\Comment{update of \texttt{N}}

	\EndWhile

	\State{$\mathtt{U}=[\mathtt{V}[j]:j\in\mathtt{H}]$} \Comment{\texttt{U}
lists the elements of $\sg(\bar{a})$ in the order they appeared in
\texttt{V}}

	\State{\Return \texttt{P}, \texttt{U}}

\EndFunction

\end{algorithmic}
\end{algorithm}

To illustrate how this algorithm functions let us work out an example.
Let $\mathbf{D}=\langle\{\top,\bot,u,u'\},\wedge,\vee\rangle$ be
the lattice from the introduction (see Figure \ref{fig rombo}). The
fundamental operations are assumed to be ordered: {[}$\wedge$, $\vee${]}.
On input $\mathbf{D}$, $\bar{a}=(u,u',\bot)$, the variables are initialized as follows:
\begin{center}
$\mathtt{V}=[u,u',\bot]$\quad{} \quad{}$\mathtt{P}=[[0],[1],[2]]$\quad{} \quad{}$\mathtt{H}=[0,1,2]$\quad{} \quad{}$\mathtt{N}=[0,1,2]$.
\par\end{center}

\noindent After the first pass through the while loop the state is:

\begin{equation*}
\begin{aligned}
\mathtt{V}=[u,u',\bot,\underbrace{u}_{u\wedge u},\underbrace{\bot}_{u\wedge u'},\underbrace{\bot}_{u\wedge\bot},\underbrace{\bot}_{u'\wedge u},\underbrace{u'}_{u'\wedge u'},\underbrace{\bot}_{u'\wedge\bot},\underbrace{\bot}_{\bot\wedge u},\underbrace{\bot}_{\bot\wedge u'},\underbrace{\bot}_{\bot\wedge\bot},& \underbrace{u}_{u\vee u},\underbrace{\top}_{u\vee u'},\underbrace{u}_{u\vee\bot}, \underbrace{\top}_{u'\vee u},\\
& \underbrace{u'}_{u'\vee u'},\underbrace{u'}_{u'\vee\bot},\underbrace{u}_{\bot\vee u},\underbrace{u'}_{\bot\vee u'},\underbrace{\bot}_{\bot\vee\bot}]
\end{aligned}
\end{equation*}

$\mathtt{P}=[[0,3,12,14,18],[1,7,16,17,19],[2,4,5,6,8,9,10,11,20],[13,15]]$

$\mathtt{H}=[0,1,2,13]$

$\mathtt{N}=[13]$.

\noindent Now the element $\top$, which was not in $\mathtt{V}$
before, makes its first appearance in position $13$. Thus, on the
next pass the fundamental operations will be applied to all pairs
with elements from $[u,u',\top,\bot]$ in which $\top$ appears at
least once. So, the next round produces:

$\mathtt{V}=\mathtt{V}+[\underbrace{u}_{u\wedge\top},\underbrace{u'}_{u'\wedge\top},\underbrace{\bot}_{\bot\wedge\top},\underbrace{u}_{\top\wedge u},\underbrace{u'}_{\top\wedge u'},\underbrace{\bot}_{\top\wedge\bot},\underbrace{\top}_{\top\wedge\top},\underbrace{\top}_{u\vee\top},\underbrace{\top}_{u'\vee\top},\underbrace{\top}_{\bot\vee\top},\underbrace{\top}_{\top\vee u},\underbrace{\top}_{\top\vee u'},\underbrace{\top}_{\top\vee\bot},\underbrace{\top}_{\top\vee\top}]$

$\mathtt{P}=[[0,3,12,14,18,21,24],[1,7,16,17,19,22,25],[2,4,5,6,8,9,10,11,20,23,26,],$

$\hspace{2.22em}[13,15,27,28,29,30,31,32,33,34]]$

$\mathtt{H}=[0,1,2,13]$

$\mathtt{N}=[\,]$.

\noindent The variable $\mathtt{N}$ is now empty due to the fact
that no new elements have been produced, so Algorithm \ref{alg:hit} halts
returning the partition $\mathtt{P}$ and $\mathtt{\mathtt{U}}=[u,u',\bot,\top]=\sg(\bar{a})$.
Running the algorithm for the tuple $\bar{b}=(u',u,\bot)$ we get:

\begin{equation*}
\begin{aligned}
\mathtt{V}=[u',u,\bot,u',\bot,\bot,\bot,u,\bot,\bot,\bot,\bot,u',\top,u',\top,u,u,u',u,\bot,u',u,\bot,& u', u, \bot,\top,\top,\\ &
\top, \top,\top,\top,\top,\top]
\end{aligned}
\end{equation*}

$\mathtt{H}=[0,1,2,13]$

$\mathtt{P}=[[0,3,12,14,18,21,24],[1,7,16,17,19,22,25],[2,4,5,6,8,9,10,11,20,23,26,],$

$\hspace{2.22em}[13,15,27,28,29,30,31,32,33,34]]$

$\mathtt{\mathtt{U}}=[u',u,\bot,\top]$.

\noindent Since the final partitions are the same for $\bar{a}$ and $\bar{b}$, by Corollary \ref{correccion_hit} below, the map $\gamma$ from $[u,u',\bot,\top]$ to $[u',u,\bot,\top]$
is a subisomorphism satisfying $\gamma:\bar{a}\mapsto\bar{b}$.

\subsection{Soundness and completeness}

For $\mathtt{X}$ a variable name in $\{\mathtt{\mathtt{V}},\mathtt{\mathtt{V}'},\mathtt{P},\mathtt{H},\mathtt{N}\}$
and $k\in\{0,\ldots,\mathrm{K}_{\bar{a}}\}$, let
\begin{itemize}
\item $\mathtt{X}_{\bar{a},k}$ be the value of variable $\mathtt{X}$ in
a run of Algorithm \ref{alg:hit} with input $\mathbf{A},\bar{a}$,
after $k$ executions of the while loop.
\end{itemize}
Also, given an arity $r$ of a function symbol of $\mathbf{A}$, let
\begin{itemize}
\item $\mathtt{T}_{\bar{a},k}^{r}$ be the value assigned to variable $\mathtt{T}$
in a run of Algorithm \ref{alg:hit} with input $\mathbf{A},\bar{a}$,
once an arity $r$ has been fixed in the for loop (line \ref{fixr}),
during the $k$th execution of the while loop.
\end{itemize}
We write $\mathtt{X}_{\bar{a}}$ for $\mathtt{X}_{\bar{a},\mathrm{K}_{\bar{a}}}$,
and for $k>\mathrm{K}_{\bar{a}}$ we define $\mathtt{X}_{\bar{a},k}$
to be $\mathtt{X}_{\bar{a}}$. The subindex $\bar{a}$
is omitted when no confusion may arise.

Given a tuple $\bar{a}\in A^{n}$ and $k\in\omega$ let $\rho_{\bar{a},k}$
be the equivalence relation over $\mathcal{T}_{k}(x_{0},\ldots,x_{n-1})$
defined by $t\mathrel{\rho_{\bar{a},k}}s\Longleftrightarrow t^{\mathbf{A}}(\bar{a})=s^{\mathbf{A}}(\bar{a})$.
Also, $Eq(Y)$ denotes the set of all equivalence relations over the
set $Y$. The following lemma provides a series of technical results
necessary to establish the soundness of Algorithm \ref{alg:hit},
which is done in Corollary \ref{correccion_hit} below.

\begin{lemma}

\label{lema tech}For any $\bar{a}\in A^{n}$ and $k\in\mathbb{\omega}$
we have:
\begin{enumerate}
\item \label{v'tieneterminos}Each member of $\mathtt{V}'_{k}$ is a term
in the variables $x_{0},\ldots,x_{n-1}$ of degree at most $k$.
\item \label{v y v' coordinados}The invariant $\mathtt{\mathtt{V}}_{k}[i]=\mathtt{V}'_{k}[i]^{\mathbf{A}}(\bar{a})$
holds throughout the for loop in line \ref{fixtup}.
\item \label{isok implica variables iguales}If $\bar{a}\approx_{k}\bar{b}$
then $\mathtt{X}_{\bar{a},k}=\mathtt{X}_{\bar{b},k}$ for $\mathtt{X}$
any variable name in $\{\mathtt{V}',\mathtt{P},\mathtt{H},\mathtt{N}\}$.
\item \label{dependencia funcional}For all $1\leq k\leq\mathrm{K}_{\bar{a}}$
there is a function 
\[
F_{k}:\mathcal{T}_{k}(x_{0},\ldots,x_{n-1})\times Eq(\mathcal{T}_{k-1}(x_{0},\ldots,x_{n-1}))\rightarrow\omega,
\]
such that for each $t\in\mathcal{T}_{k}(x_{0},\ldots,x_{n-1})$ the
index $i=F_{k}(t,\rho_{\bar{a},k-1})$ satisfies $i\leq\left|\mathtt{\mathtt{V}}_{k}\right|-1$
and $t^{\mathbf{A}}(\bar{a})=\mathtt{\mathtt{V}}_{k}[i]$.
\item \label{Pk determina anteriores}For all $k\leq\mathrm{K}_{\bar{a}}$
we have that $\mathtt{P}_{k}$ determines $\mathtt{X}_{j}$ for all
$j<k$ and $\mathtt{X}$ any variable name in $\{\mathtt{V}',\mathtt{P},\mathtt{H},\mathtt{N}\}$.
\item \label{P determina iso}If $\mathtt{P}_{\bar{a}}=\mathtt{P}_{\bar{b}}$,
then $\bar{a}\approx\bar{b}$.
\end{enumerate}
\end{lemma}

\begin{proof}(\ref{v'tieneterminos}) is easily proved by induction
in $k$. (\ref{v y v' coordinados}) is clear from lines \ref{beginvv'}
to \ref{endvv'}. 

\noindent \medskip{}

\noindent (\ref{isok implica variables iguales}) We proceed by induction in $k$. The case $k=0$
is straightforward. Assume now $k+1\leq\min(\mathrm{K}_{\bar{a}},\mathrm{K}_{\bar{b}})$,
and suppose $\bar{a}\approx_{k+1}\bar{b}$. Since $\bar{a}\approx_{k}\bar{b}$,
by our inductive hypothesis we have that
\begin{itemize}
\item $\mathtt{X}_{\bar{a},k}=\mathtt{X}_{\bar{b},k}$ for $\mathtt{X}$
any variable name in $\{\mathtt{V}',\mathtt{P},\mathtt{H},\mathtt{N}\}$.
\end{itemize}
Note that the first line executed when starting the $k+1$th run of
the while loop sets the value of $\mathtt{H\_old}$ to the value
of $\mathtt{H}_{k}$. Now, observe that the value assigned to $\mathtt{T}$
for each arity $r$ only depends on $\mathtt{H\_old}$ and $\mathtt{N}_{k}$.
Thus $\mathtt{T}_{\bar{a},k+1}^{r}=\mathtt{T}_{\bar{b},k+1}^{r}$
for all arities $r$. This, along with the fact that the function
symbols at the for loop starting at Line \ref{fixf} appear always in
the same order, implies that $\mathtt{V}'_{\bar{a},k+1}=\mathtt{V}'_{\bar{b},k+1}$.
We prove next that $\mathtt{P}_{\bar{a},k+1}=\mathtt{P}_{\bar{b},k+1}$;
suppose the indexes $i$ and $j$ are in the same block of $\mathtt{P}_{\bar{a},k+1}$.
That is, $\mathtt{\mathtt{V}}_{k+1}(\bar{a})[i]=\mathtt{\mathtt{V}}{}_{k+1}(\bar{a})[j]$,
so, by (\ref{v y v' coordinados}), we have $\mathtt{V}'_{\bar{a},k+1}[i]^{\mathbf{A}}(\bar{a})=\mathtt{V}'_{\bar{a},k+1}[j]^{\mathbf{A}}(\bar{a})$.
Now, (\ref{v'tieneterminos}) says that $\mathtt{V}'_{k+1}(\bar{a})[i]$
and $\mathtt{V}'_{k+1}(\bar{a})[j]$ are terms of degree at most $k+1$,
and they agree on $\bar{a}$. Hence they agree on $\bar{b}$, since
$\bar{a}\approx_{k+1}\bar{b}$. Invoking (\ref{v y v' coordinados})
once again yields that $i$ and $j$ are in the same block of $\mathtt{P}_{\bar{b},k+1}$.
From the fact that $\mathtt{P}_{\bar{a},k+1}=\mathtt{P}_{\bar{b},k+1}$
it easy to see that $\mathtt{H}_{\bar{a},k+1}=\mathtt{H}_{\bar{b},k+1}$
and $\mathtt{N}_{\bar{a},k+1}=\mathtt{N}_{\bar{b},k+1}$. In particular,
for any $l\leq\min(\mathrm{K}_{\bar{a}},\mathrm{K}_{\bar{b}})$ we
have that $\mathtt{N}_{\bar{a},l}=\emptyset$ if and only if $\mathtt{N}_{\bar{b},l}=\emptyset$.
So $\mathrm{K}_{\bar{a}}=\mathrm{K}_{\bar{b}}$, from which it follows
at once that (\ref{isok implica variables iguales}) holds for $k+1>\min(\mathrm{K}_{\bar{a}},\mathrm{K}_{\bar{b}})$.\medskip{}

\noindent (\ref{dependencia funcional}) Take $t=f(t_{0},\ldots,t_{r-1})\in\mathcal{T}_{k+1}$
and suppose $k+1\leq\mathrm{K}_{\bar{a}}$. For $k=0$, define $F_{0}(x_{j})=j$.
If $k\geq1$, by inductive hypothesis we have the function $F_{k}$,
and thus we can obtain indexes $i_{0}:=F_{k}(t_{0},\rho_{k-1}),\ldots,i_{r-1}:=F_{k}(t_{r-1},\rho_{k-1})$
functionally from $t$ and $\rho_{k}$ (obviously $t_{0},\ldots,t_{r-1}$
can be obtained as functions of $t$ and $\rho_{k-1}=\rho_{k}\cap\mathcal{T}_{k-1}^{2}$).
If $k=0$, just take $i_{0}:=F_{0}(t_{0}),\ldots,i_{r-1}:=F_{0}(t_{r-1})$.
Next, define

\begin{align*}
\tilde{t}_{0} & :=\mathtt{V}'_{k}[i_{0}]\\
 & \vdots\\
\tilde{t}_{r-1} & :=\mathtt{V}'_{k}[i_{r-1}],
\end{align*}
and once again, note that these $\tilde{t}_{l}$ can be obtained functionally
since $\mathtt{V}'_{k}$ can be obtained functionally from $\rho_{k}$.
Also, by our inductive hypothesis, we have $t^{\mathbf{A}}(\bar{a})=f(\tilde{t}_{0},\ldots,\tilde{t}_{r-1})^{\mathbf{A}}(\bar{a})$.

Assume first that there is $l\in\{0,\ldots,r-1\}$ such that $i_{l}\in\mathtt{H}_{k}\setminus\mathtt{H}_{k-1}$.
Then $i_{l}\in\mathtt{N}_{k}$, so $\text{\ensuremath{\left\langle i_{0},\ldots,i_{r-1}\right\rangle }}\in\mathtt{T}_{k+1}^{r}$,
and we have that the term $f(\tilde{t}_{0},\ldots,\tilde{t}_{r-1})$
is appended to $\mathtt{V}'$, say with index $j$, during the $(k+1)$th
pass of the while loop. Observe that $j$ only depends on the signature
of $\mathbf{A}$ and the numbers $\left|\mathtt{T}_{k+1}^{u}\right|$
for $u$ an arity less or equal than $r$. But $\mathtt{T}_{k+1}^{u}$
depends only on $\mathtt{H}_{k}$ and $\mathtt{N}_{k}$, which in
turn are determined by $\rho_{k}$. Thus $j$ can be obtained as a
function of $t$ and $\rho_{k}$, and we can define $F_{k+1}(t,\rho_{k}):=j$
in the case that there is $l\in\{0,\ldots,r-1\}$ such that $i_{l}\in\mathtt{H}_{k}\setminus\mathtt{H}_{k-1}$.
Suppose on the other hand that $i_{0},\ldots,i_{r-1}\in\mathtt{H}{}_{k-1}$.
Then, by (\ref{v'tieneterminos}), we have that $\tilde{t}_{0},\ldots,\tilde{t}_{r-1}\in\mathcal{T}_{k-1}$,
and thus $f(\tilde{t}_{0},\ldots,\tilde{t}_{r-1})\in\mathcal{T}_{k}$.
Hence, in this case we can define $F_{k+1}(t,\rho_{k}):=F_{k}(f(\tilde{t}_{0},\ldots,\tilde{t}_{r-1}),\rho_{k-1})$.\medskip{}

\noindent (\ref{Pk determina anteriores}) Notice first that, for
$j\leq k$, we have $\mathtt{P}_{j}$ is the restriction of $\mathtt{P}_{k}$
to the set $\{0,...,\left|\mathtt{\mathtt{V}}_{j}\right|-1\}$. Also,
$\mathtt{H}_{j}=[\min(B):B\in\mathtt{P}_{j}]$, so that $\mathtt{P}_{j}$
determines $\mathtt{H}_{j}$. On the other hand, $\mathtt{N}_{0}$
is defined as $\mathtt{H}_{0}$, while $\mathtt{N}_{j+1}=\mathtt{H}_{j+1}\setminus\mathtt{H}_{j}$.
Observe also that $\mathtt{V}'_{0}$ is also determined just by the
length of $\bar{a}$, and that $\mathtt{V}'_{j+1}$ is obtained from
$\mathtt{H}_{j}$ and $\mathtt{N}_{j}$, so it is determined by $\mathtt{H}_{j}$
and $\mathtt{H}_{j-1}$. Using these observations, the conclusion
follows now from an easy inductive argument.

\medskip{}

\noindent (\ref{P determina iso}) Let $K=\mathrm{K}_{\bar{a}}$.
By (\ref{Pk determina anteriores}), for every $k\leq K$ we have
$\mathtt{P}_{\bar{a},k}=\mathtt{P}_{\bar{b},k}$ and $\mathtt{V}'_{\bar{a},k}=\mathtt{V}'_{\bar{b},k}$.
Let $t,s\in\mathcal{T}_{K}$ such that $t^{\mathbf{A}}(\bar{a})=s^{\mathbf{A}}(\bar{a})$.
Define $i=F_{K}(t,\rho_{\bar{a},K-1})$ and $j=F_{K}(s,\rho_{\bar{a},K-1})$.
Define also $\hat{t}=\mathtt{V}'_{\bar{a},k}[i]$ and $\hat{s}=\mathtt{V}'_{\bar{a},k}[j]$.
This means $\hat{t}^{\mathbf{A}}(\bar{a})=\hat{s}^{\mathbf{A}}(\bar{a})$
and, since $\mathtt{V}'_{\bar{a},K}=\mathtt{V}'_{\bar{b},K}$, we
have $\hat{t}^{\mathbf{A}}(\bar{b})=\hat{s}^{\mathbf{A}}(\bar{b})$,
which in turn implies $t^{\mathbf{A}}(\bar{b})=s^{\mathbf{A}}(\bar{b})$.
This shows $\bar{a}\approx_{K}\bar{b}$, so by Lemma \ref{lema mat}.(\ref{isoK implica iso})
we have $\bar{a}\approx\bar{b}$.

\end{proof}

Given a finite algebra $\mathbf{A}$ and $\bar{a}\in A^{n}$, define
$\ity(\bar{a})$ and $\iun(\bar{a})$ by $\mathrm{isoType}(\mathbf{A},\bar{a})=(\ity(\bar{a}),\iun(\bar{a}))$.
Notice that $\iuna a$ is a list without repetitions of all the elements
in $\sg(\bar{a})$.

The next corollary summarizes the results collected in Lemma \ref{lema tech}
using the terminology just introduced. It shows that the output of
our algorithm captures the isomorphism type of a tuple $\bar{a}$ in
a finite algebraic structure.

\begin{corollary}

\label{correccion_hit}Let $\mathbf{A}$ be a finite algebra. Then for
any $\bar{a},\bar{b}\in A^{n}$ the following are equivalent:
\begin{enumerate}
\item $\bar{a}\approx\bar{b}$
\item $\ity(\bar{a})=\ity(\bar{b})$.
\end{enumerate}
Moreover, if either of these conditions hold, then the map $\gamma:\iun(\bar{a})[j]\mapsto\iun(\bar{b})[j]$
for $j\in\{0,...,|\iun(\bar{a})|-1\}$ is an isomorphism from $\mathbf{Sg}(\bar{a})$
into $\mathbf{Sg}(\bar{b})$ mapping $\bar{a}$ to $\bar{b}$.

\end{corollary}

\begin{proof}
\noindent If $\bar{a}\approx\bar{b}$, in view of Lemma \ref{lema tech}.(\ref{isok implica variables iguales})
we have $\ity(\bar{a})=\ity(\bar{b})$. Conversely, if $\ity(\bar{a})=\ity(\bar{b})$,
then by lemma (\ref{lema tech}).(\ref{P determina iso}) we have
$\bar{a}\approx\bar{b}$. In either case, by lemma (\ref{lema tech}).(\ref{isok implica variables iguales}),
we get $\mathtt{X}_{\bar{a}}=\mathtt{X}_{\bar{b}}$ for $\mathtt{X}\in\{\mathtt{V}',\mathtt{P},\mathtt{H},\mathtt{N}\}$.
In particular, $\mathtt{H}_{\bar{a}}=\mathtt{H}_{\bar{b}}$, from
which it follows that $|\iun(\bar{a})|=|\iun(\bar{b})|$. Since all
the elements in $\iuna a$ (and also those in $\iuna b$) are different,
we have that $\gamma$ is a well-defined bijection. Also, since $\mathtt{P}_{\bar{a},0}=\mathtt{P}_{\bar{b},0}$,
we have that $\gamma$ sends each $a_{j}$ into $b_{j}$. To prove
$\gamma$ is an homomorphism, let $t$ be a term. Let $i\in\mathtt{H}_{\bar{a}}$
be such that $t^{\mathbf{A}}(\bar{a})=\mathtt{\mathtt{V}}_{\bar{a}}[i]$,
and define $\hat{t}=\mathtt{V}'_{\bar{a}}[i]$. By (\ref{v y v' coordinados}),
$t$ and $\hat{t}$ agree on $\bar{a}$, hence they also agree on
$\bar{b}$. So we have $\gamma(t^{\mathbf{A}}(\bar{a}))=\gamma(\hat{t}^{\mathbf{A}}(\bar{a}))=\mathtt{\mathtt{V}}{}_{\bar{b}}[i]=\mathtt{\mathtt{V}'}_{\bar{b}}[i]^{\mathbf{A}}(\bar{b})$,
and since $\mathtt{\mathtt{V}'}_{\bar{a}}=\mathtt{V}'_{\bar{b}}$,
this equals $\hat{t}^{\mathbf{A}}(\bar{b})=t^{\mathbf{A}}(\bar{b})=t^{\mathbf{A}}(\gamma(\bar{a}))$.

\end{proof}

\section{Computing QfDefAlg with a merging strategy}\label{sec:megahit}

In this section we present the first algorithm we developed, referred to as the \emph{Merging Algorithm} in the sequel, to decide qf-definability
in a finite algebraic structure. It is based on a slight
improvement of Theorem \ref{thm:lema sem}, stated below as Corollary
\ref{cor:lema semantico sin repeticiones}. Before executing the algorithm, it is convenient to preprocess its input as explained in the following subsection.

\subsection{Preprocessing the input relation} \label{preprocesamiento}

Let $\bar{a}=(a_{0},\dots,a_{k-1})$ be a tuple. We define the \emph{pattern}
of $\bar{a}$, denoted by $\pat\bar{a}$, to be the partition of $\{0,\dots,n-1\}$
such that $i,j$ are in the same block if and only if $a_{i}=a_{j}$.
For example, $\pat(a,a,b,c,b,c)=\{\{0,1\},\{2,4\},\{3,5\}\}$. We
write $\left\lfloor \bar{a}\right\rfloor $ to denote the tuple obtained
from $\bar{a}$ by deleting every entry equal to a prior entry. E.g.,
$\left\lfloor (a,a,b,c,b,c)\right\rfloor =(a,b,c)$. Let $R$ be a
relation and $\theta$ a pattern, we define:
\begin{itemize}
\item $\left\lfloor R\right\rfloor :=\{\left\lfloor \bar{a}\right\rfloor :\bar{a}\in R\}$,
\item $\spec(R):=\{\left|\left\lfloor \bar{a}\right\rfloor \right|:\bar{a}\in R\}$,
\item $R_{\theta}:=\{\bar{a}\in R:\pat\bar{a}=\theta\}$
\end{itemize}
Given a tuple $\bar{a}$ from $A$ and relations $R_{0},\ldots,R_{l-1}$
over $A$, let $\rty(\bar{a},R_{0},\ldots,R_{l-1})\in\{\mathrm{True},\mathrm{False}\}^{l}$
be the tuple $(R_{0}(\bar{a}),\ldots,R_{l-1}(\bar{a}))$. For a positive
integer $k$ let $A^{(k)}:=\{(a_{0},\dots,a_{k-1})\in A^{k}:a_{i}=a_{j}\Leftrightarrow i=j\}$.

\begin{corollary}
\label{cor:lema semantico sin repeticiones}Let $\mathbf{A}$ be a
finite algebra, $R\subseteq A^{m}$ and let $R_{0},\ldots,R_{l-1}$
be all the relations of the form $\left\lfloor R_{\pat\bar{a}}\right\rfloor $
for $\bar{a}\in R$. The following are equivalent:
\begin{enumerate}
    \item $R$ is qf-definable in $\mathbf{A}$. \label{r_def_A}
    \item $R_{0},\ldots,R_{l-1}$ are qf-definable in $\mathbf{A}$. \label{r_def_B}
    \item For all $k\in\spec(R)$ and for all $\bar{a},\bar{b}\in A^{(k)}$
we have that $\bar{a}\approx\bar{b}$ implies $\rty(\bar{a},R_{0},\ldots,R_{l-1})=\rty(\bar{b},R_{0},\ldots,R_{l-1})$. \label{r_def_C}       
\end{enumerate}
\end{corollary}
\begin{proof} The equivalence (\ref{r_def_A}) $\Leftrightarrow$ (\ref{r_def_B}) is an easy exercise,
and (\ref{r_def_B}) $\Leftrightarrow$ (\ref{r_def_C}) is just a restatement of Theorem \ref{thm:lema sem}.
\end{proof}

So, before executing our algorithms we can strip down the target input relation into relations without superfluous information. This can have a serious impact on performance, given
that the search space depends exponentially on the arity of the target. To do this, in view of the above corollary, we just need to compute all the relations of the form $\left\lfloor  R_{\pat\bar{a}}\right\rfloor $ for $\bar{a}\in R$. 

\subsection{Breakdown of the Merging Algorithm}

Let $\mathbf{A}$ be an finite algebra and let $R\subseteq A^{n}$.
The merging strategy to decide if $R$ is qf-definable in $\mathbf{A}$
can be summarized as follows.
\begin{enumerate}
\item Preprocess the input relation as explained in subsection \ref{preprocesamiento}, obtaining relations $R_{0},\ldots,R_{l-1}$.
\item For each $k\in\spec(R)$ compute the partition induced by the equivalence
relation $\approx$ on $A^{(k)}$.
\item If for some $k\in\spec(R)$ there are $\bar{a},\bar{b}\in A^{(k)}$
such that $\rty(\bar{a},R_{0},\ldots,R_{l-1})\neq\rty(\bar{b},R_{0},\ldots,R_{l-1})$ and $\bar{a}\approx\bar{b}$
return $\false$. Otherwise, return $\true$.
\end{enumerate}

An obvious method for carrying out steps (2) and (3) is to use the
function isoType to tag each tuple in $\bigcup_{k\in\spec(R)}A^{(k)}$
with its isomorphism type, and every time two tuples have the same
iType check that they have the same relType with respect to $R_{0},\ldots,R_{l-1}$.
Clearly, it is convenient to carry out this test as soon as two tuples
with the same isomorphism type are discovered, because if it fails
the algorithm can stop (and return $\false$). Another
improvement is that, whenever we find tuples $\bar{a},\bar{b}$
with the same isomorphism type, we have access (essentially without
any further computational cost) to an isomorphism $\gamma$ between
$\sg(\bar{a})$ and $\sg(\bar{b})$. If two tuples are connected by
$\gamma$, they have the same isomorphism type, and thus it suffices
to compute the isomorphism type of one of them. These observations
together with a deliberate strategy to cycle through the tuples comprise
the main ideas behind our algorithm. We explain next how they are
implemented. The pseudocode for the Merging Algorithm is exhibited in Algorithm \ref{alg:open} with its crucial subroutine displayed in Algorithm \ref{alg:propagar}.  

At every point during an execution, the data stored in $\mathtt{orbits}_{k}$
is a partition of $A^{(k)}$ where each block is annotated with some
additional information. We call these annotated blocks \emph{orbits},
and they have the form $(B,RT,T,U)$ where $B\subseteq A^{(k)}$ is
the actual block, $RT$ is the relType of all tuples in $B$, $T$
shall store the $\mathrm{iType}$ of the tuples in $B$ (when known),
and $U$ shall store the $\mathrm{iUniv}$ of a tuple in $B$ (when
known). When the algorithm starts, every block is a singleton annotated
with its relType, and $T$,$U$ are set to Null.
The algorithm traverses and processes the tuples in $\bigcup_{k\in\spec(R)}A^{(k)}$
in a DFS-like fashion. The \emph{tree }that is traversed is comprised
by the subuniverses of $\mathbf{A}$ and has $A$ as its root. At
every point of an execution, the algorithm is working on a node $S$
of this tree by processing the tuples in $\bigcup_{k\in\spec(R)}S^{(k)}$.
To keep track of the current and previously visited nodes, and the
tuples already processed at each node, a stack is used. The entries
in the stack have the form $(S,L,G)$ where $S$ is a subuniverse,
$L$ is the list of tuples from $S$ still to be processed, and $G\subseteq\bigcup_{k\in\spec(R)}S^{(k)}$
is a set of tuples such that: every tuple in $G$ generates $S$,
and no two tuples in $G$ have the same isomorphism type. (The point
of keeping track of such a $G$ shall become clear below.) The stack
is initialized with $(A,L_{0},\emptyset)$ where $L_{0}$ is a list
of the tuples in $\bigcup_{k\in\spec(R)}A^{(k)}$ ordered increasingly
by arity. The first tuple processed during an execution is the first
element in $L_{0}$, and every time the algorithm is ready to process
a new tuple, there is a triplet $(S,L,G)$ on the top of the stack
and the first element of $L$ is popped and processed. To see how
a tuple is processed, suppose the top of the stack is $(S,L,G)$ and
a tuple $\bar{a}$ has been popped from $L$. If the isomorphism type
of $\bar{a}$ is known (i.e., if $\bar{a}$ belongs to an orbit with
known type), we pop the next tuple in $L$ (if there are no more tuples
in $L$, the entry $(S,L,G)$ is removed from the stack). If, on the
other hand, the type of $\bar{a}$ is unknown, the function isoType
is called to compute it, and the next step is to search through the
orbits to see if there is one tagged with $\itya a$. However, which
orbits we inspect (and the behaviour of the algorithm afterwards)
depends on whether $\sg(\bar{a})$ is smaller than $S$.

\begin{algorithm}
\caption{}
\label{alg:open}

\begin{algorithmic}[1]

\Function{\texttt{$\mathtt{MergingAlgorithm}$}}{$\mathbf{A}$, $R$}\Comment{$\mathbf{A}$ is an algebra and \texttt{$R$} is a relation
over $A$}

	\State{$\mathtt{targets}=(R_{1},\ldots,R_{l})$ $R_{j}$'s
are all the distinct relations of the form $\left\lfloor R_{\pat\bar{a}}\right\rfloor $
for $\bar{a}\in R$.}

	\State{$\mathtt{spec}=\mathtt{sorted}(\spec(R))$}\Comment{sorted
increasingly}

	\For{$k\in\mathtt{spec}$}

		\State{$\mathtt{orbits}_{k}=\{(\{\bar{a}\},\text{relType}(\bar{a},\mathtt{targets}),\unk,\unk):\bar{a}\in A^{(k)}\}$}\Comment{initialization}

	\EndFor

	\State{$\mathtt{orbits}=\{\mathtt{orbits}_{k}:k\in\mathtt{spec}\}$}

	\State{$\mathtt{tuples\_to\_process}=\mathtt{sorted}(\bigcup_{k\in\mathtt{spec}}A^{(k)})$}\Comment{sorted
decreasingly by arity}

	\State{$\mathtt{stack}=[(A,\mathtt{tuples\_to\_process},\emptyset)]$}\Comment{initialization
of the stack}

	\While{$\mathtt{stack}$ is not empty}

		\State{Let $(\text{\texttt{current\_sub}},\text{\texttt{tuples\_to\_process}},\text{\texttt{generators}})$ reference the top of \texttt{stack}}

		\While{$\text{\texttt{tuples\_to\_process}}\neq[\ ]$}

			\State{$\bar{a}=\text{pop first element of }\mathtt{tuples\_to\_process}$}

			\If{$\type(\bar{a})=\unk$}\Comment{$\bar{a}$'s type is unknown}

				\State{$(\mathtt{type\_a},\mathtt{universe\_a})=\mathtt{IsoType}(\boldsymbol{A},\bar{a})$}

				\If{$|\mathtt{universe\_a}|=|\mathtt{current\_sub}|$}\Comment{$\sg(\bar{a})$
is not smaller}

					\If{$\exists$ $(\mathtt{B},\mathtt{RT},\mathtt{T},\mathtt{U})\in\{\text{orbits in }\mathtt{generators}\}$
such that $\mathtt{type\_a}=\mathtt{T}$ }

						\State{$\gamma=$ the isomorphism from $\mathtt{universe\_a}$
to $\mathtt{U}$}

						\If{\texttt{not }$\mathtt{try\_merge\_orbits}(\gamma,\mathtt{orbits})$}\Comment{see
Algorithm \ref{alg:propagar}}

							\State{\Return $\false$}
						\EndIf

					\Else\Comment{$\bar{a}$'s type is new}

						\State{$\mathtt{tag\_orbit}(\bar{a},\mathtt{type\_a},\mathtt{universe\_a})$}\label{tag_orbit no achica}

						\State{$\text{add }\bar{a}\text{ to }\mathtt{generators}$}

					\EndIf

				\Else\Comment{$\sg(\bar{a})$ is smaller}

					\If{there is $(\mathtt{B},\mathtt{RT},\mathtt{T},\mathtt{U})\in\mathtt{orbits}_{|\bar{a}|}$
such that $\mathtt{type\_a}=\mathtt{T}$}\Comment{not new}

						\State{$\gamma=$ the isomorphism from $\mathtt{universe\_a}$
to $\mathtt{U}$}

						\If{\texttt{not }$\mathtt{try\_merge\_orbits}(\gamma,\mathtt{orbits})$}\Comment{see
Algorithm \ref{alg:propagar}}

							\State{\Return $\false$}

						\EndIf

					\Else\Comment{$\bar{a}$'s type is new}

						\State{$\mathtt{tag\_orbit}(\bar{a},\mathtt{type\_a},\mathtt{universe\_a})$}\Comment{then push new node on \texttt{stack}}

						\State{push $(\mathtt{universe\_a},\mathtt{sorted}(\bigcup_{k\in\mathtt{spec}}\mathtt{universe\_a}^{(k)}),\{\bar{a}\})$ %onto \texttt{stack}
}
						\State{$\mathtt{tag\_orbit}(\bar{a},\mathtt{type},\mathtt{universe})$}\label{tag_orbit achica}

						\State{$\breakk$}

					\EndIf

				\EndIf

			\EndIf

		\EndWhile

		\If{$\mathtt{tuples\_to\_process}=[\ ]$}

			\State{delete top of the \texttt{stack}}

		\EndIf

	\EndWhile

\State{\Return $\true$}

\EndFunction

\end{algorithmic}
\end{algorithm}

\noindent \textbf{Case }$\left|\sg(\bar{a})\right|=\left|S\right|$.
Here we only inspect the orbits of tuples in $G$ for one tagged with
$\itya a$ (to see why this suffices see Lemma \ref{propiedades del stack}).
If none of these orbits is tagged with $\itya a$, we call the function
\texttt{tag\_orbit} to tag $\bar{a}$'s
orbit with $\itya a$ and $\iuna a$, and add $\bar{a}$ to $G$.
If, on the other hand, there is an orbit $(B,RT,T,U)$ of an element
in $G$ with $T=\itya a$, then, by Corollary \ref{correccion_hit}, the
function $\gamma:\iuna a[i]\mapsto U[i]$ is a subisomorphism of $\mathbf{A}$.
So, we proceed to merge orbits containing tuples connected by $\gamma$.
In particular, the orbit containing $\bar{a}$ is merged with $(B,RT,T,U)$,
and thus $\bar{\ensuremath{a}}$ ends up in an orbit tagged with $\itya a$.
The merging is done by the function \texttt{try\_merge\_orbits} (see Algorithm \ref{alg:propagar}), which
also checks that any two orbits to be merged have the same relType
(if this test fails, Algorithm \ref{alg:open} stops and returns $\false$).

\noindent \textbf{Case }$\left|\sg(\bar{a})\right|<\left|S\right|$.
In this case we inspect all orbits in $\mathtt{orbits}_{\left|\bar{a}\right|}$
in search for one with isomorphism type $\itya a$. If there is no
such orbit, then orbit containing $\bar{a}$ is annotated with $\itya a$
and $\iuna a$. We also push $(\sg(\bar{a}),L',\{\bar{a}\})$ on top
of the stack (i.e. we move to a new node of the subuniverse tree).
If there is an orbit in $\mathtt{orbits}_{\left|\bar{a}\right|}$
tagged with $\itya a$, we proceed in the same way as above by calling
\texttt{try\_merge\_orbits}.

\noindent Eventually, if no call to \texttt{try\_merge\_orbits} returns
$\false$, every tuple in $\bigcup_{k\in\spec(R)}A^{(k)}$ is processed,
and when the algorithm halts we have computed for each $k\in\spec(R)$
the partition induced by $\approx$ on $A^{(k)}$. Since we ensure
that all tuples in the same orbit have the same relType, Corollary
\ref{cor:lema semantico sin repeticiones} guarantees that $R$ is
qf-definable in $\mathbf{A}$.

\begin{algorithm}[H]
\caption{}
\label{alg:propagar}

\begin{algorithmic}[1]

\Function{\texttt{$\mathtt{try\_merge\_orbits}$}}{$\gamma$, \texttt{orbits}}

	\For{$k\in\mathtt{spec}$}

		\For{$\bar{a}\in(Dom(\gamma))^{(k)}$}

			\If{$\mathtt{orbit}(\bar{a})\neq\mathtt{orbit}(\gamma(\bar{a}))$}

				\State{$(\mathtt{B},\mathtt{RT},\mathtt{T},\mathtt{U})=\mathtt{orbit}(\bar{a})$}

				\State{$(\mathtt{B'},\mathtt{RT'},\mathtt{T'},\mathtt{U'})=\mathtt{orbit}(\gamma(\bar{a}))$}

				\If{$\mathtt{RT}\neq\mathtt{RT'}$}

					\State{\Return $\false$}\Comment{RelTypes differ}

				\EndIf

				\State{$\mathtt{B\_merge}=\mathtt{B}\cup\mathtt{B'}$}\Comment{blocks
are merged}

				\State{$\mathtt{RT\_merge}=\mathtt{RT}$}

				\If{$\mathtt{T}\neq\unk$}\Comment{first block was tagged}

					\State{$\mathtt{T\_merge}=\mathtt{T}$}

					\State{$\mathtt{U\_merge}=\mathtt{U}$}

				\ElsIf{$\mathtt{T'}\neq\unk$}\Comment{second block was tagged}

					\State{$\mathtt{T\_merge}=\mathtt{T'}$}

					\State{$\mathtt{U\_merge}=\mathtt{U'}$}

				\Else\Comment{no block is tagged}

					\State{$\mathtt{T\_merge}=\unk$}

					\State{$\mathtt{U\_merge}=\unk$}

				\EndIf

				\State{delete $(\mathtt{B},\mathtt{RT},\mathtt{T},\mathtt{U})$
and $(\mathtt{B'},\mathtt{RT'},\mathtt{T'},\mathtt{U'})$ from $\mathtt{orbits}_{k}$}

				\State{add $(\mathtt{B\_merge},\mathtt{RT\_merge},\mathtt{T\_merge},\mathtt{U\_merge})$
to $\mathtt{orbits}_{k}]$}

			\EndIf

		\EndFor

	\EndFor

\State{\Return$\true$}

\EndFunction

\end{algorithmic}
\end{algorithm}

\subsection{Proof of soundness and completeness}

To prove soundness and completeness we need some auxiliary lemmas.

\begin{lemma}

\label{lem:invariantes orbitas}The following properties hold throughout
an execution of Algorithm \ref{alg:open} for each $k\in\spec(R)$:
\begin{enumerate}
    \item \label{particion}The set $\{B:(B,RT,T,U)\in\mathtt{orbits}_{k}\}$
is a partition of $A^{(k)}$.
    \item \label{tipo y universo}For each $(B,RT,T,U)\in\mathtt{orbits}_{k}$
we have that:
    \begin{enumerate}
        \item \label{relType correcto}For all $\bar{a}$ in $B$ we have $\rty(\bar{a},R_{0},\ldots,R_{l-1})=RT$.
        \item For all $\bar{a},\bar{b}$ in $B$ we have $\bar{a}\approx\bar{b}$.
        \item If $U\neq\unk$, then there is $\bar{a}_{0}\in B$ such that $\iun(\bar{a}_{0})=U.$
        \item If $T\neq\unk$, then $T=\itya a$ for every $\bar{a}$ in $B$.
    \end{enumerate}
\end{enumerate}

\end{lemma}

\begin{proof}
The properties are obviously true at the initialization of $\mathtt{orbits}_{k}$.
Note that the only instructions of Algorithm \ref{alg:open} that
may change $\mathtt{orbits}_{k}$ are calls to the functions \texttt{try\_merge\_orbits}
and \texttt{tag\_orbit}, and it is easy to see that these functions
preserve (\ref{particion}) and (\ref{tipo y universo}).
\end{proof}

In order to make the proofs below more succinct, we say that (at a
certain point of an execution of Algorithm \ref{alg:open}):
\begin{itemize}
\item an orbit $(B,RT,T,U)$ is \emph{tagged }if $T\neq\unk$.
\item the tuple $\bar{a}$ has \emph{known type }if the (unique) orbit containing
$\bar{a}$ is tagged.
\end{itemize}

\begin{lemma}
\label{propiedades del stack}The following properties hold throughout
an execution of Algorithm \ref{alg:open}:
\begin{enumerate}
\item \label{stack decreciente}If the stack is $[(S_{m},L_{m},G_{m}),(S_{m-1},L_{m-1},G_{m-1}),\ldots,(S_{0},L_{0},G_{0})]$
then $\left|S_{m}\right|<\dots<\left|S_{0}\right|$.
\item \label{todas etiquetadas}If an entry $(S,L,G)$ is removed from the
stack, then every tuple in $\bigcup_{k\in\spec(R)}S^{(k)}$ has known
type.
\item \label{si achica todas etiquetadas}Suppose the top of the stack is
$(S,L,G)$. If $\bar{a}$ is a tuple with known type such that $\left|\sg(\bar{a})\right|<\left|S\right|$,
then every tuple in $\bigcup_{k\in\spec(R)}\sg(\bar{a})^{(k)}$ has
known type.
\item \label{si no achica y no esta conectada todas etiquetadas}Suppose
the top of the stack is $(S,L,G)$. If $\bar{a}$ is a tuple with
known type such that $\left|\sg(\bar{a})\right|=\left|S\right|$,
then either:
\begin{enumerate}
\item there is $\bar{g}\in G$ and $\bar{g}$ is in the same orbit as $\bar{a}$,
or
\item every tuple in $\bigcup_{k\in\spec(R)}\sg(\bar{a})^{(k)}$ has known
type.
\end{enumerate}
\item \label{etiquetas unicas}For all $k\in\spec(R)$, and for any $O=(B,RT,T,U)$
and $O'=(B',RT',T',U')$ in $\mathtt{orbits}_{k}$ such that $T\neq\unk$,
we have that 
\[
T=T'\Longleftrightarrow O=O'.
\]
\end{enumerate}
\end{lemma}

\begin{proof}
Item (\ref{stack decreciente}) is easily seen to hold. We prove (\ref{todas etiquetadas}).
When an element $(S,L,G)$ is pushed onto the stack then $L$ is a
list containing all tuples in $\bigcup_{k\in\spec(R)}S^{(k)}$, and
$(S,L,G)$ is only removed from the stack when all tuples in $T$
have been processed. It is not hard to see that when a tuple $\bar{a}$
is processed, we find that either:
\begin{itemize}
\item the orbit containing it is already tagged , or
\item it is tagged by a call to \texttt{tag\_orbit} or \texttt{try\_merge\_orbits},
unless the call to \texttt{try\_merge\_orbits} returns $\false$ (making
Algorithm \ref{alg:open} halt and return $\false$).
\end{itemize}
Let us prove (\ref{si achica todas etiquetadas}). Suppose the top
of the stack is $(S,L,G)$, and let $\bar{a}$ be as in the statement.
Note that there are two ways in which the orbit containing $\bar{a}$
could have been tagged: either by a call to \texttt{tag\_orbit} or
by the merging of $\bar{a}$'s orbit with another orbit that was already
tagged. Suppose first that at some point the algorithm executed the
call $\mathtt{tag\_orbit}(\bar{a},\itya a,\iuna a)$. The key observation
is that then $(\sg(\bar{a}),L',G')$ has been on the stack. Now, since
$\left|\sg(\bar{a})\right|<\left|S\right|$, item (\ref{stack decreciente})
says that $(\sg(\bar{a}),L',G')$ is no longer on the stack, and,
by (\ref{todas etiquetadas}), it follows that every tuple
from $\sg(\bar{a})$ has known type. Next, assume that $\bar{a}$'s
orbit was not tagged by such a call to \texttt{tag\_orbit.}Since the
only way to introduce a new type in $\mathtt{orbits}$ is with a call
to \texttt{tag\_orbit}, there must exist a tuple $\bar{b}$ such that:
\begin{itemize}
\item $\itya b=\itya a$, and $\mathtt{tag\_orbit}(\bar{b},\itya a,\iuna b)$
has been executed at an earlier time, and
\item $\bar{a}$'s orbit was tagged by eventually being merged with $\bar{b}$'s
orbit.
\end{itemize}
Using the same reasoning as above we can conclude that every tuple
from $\sg(\bar{b})$ has known type. Finally, observe that through
the merging process every tuple from $\sg(\bar{a})$ ends up in an
orbit with a tuple from $\sg(\bar{b})$, and thus has known type.

We leave (\ref{si no achica y no esta conectada todas etiquetadas})
to the reader as the proof is similar to the one of (\ref{si achica todas etiquetadas}).

To conclude we prove (\ref{etiquetas unicas}). Note that it suffices
to check that the calls to \texttt{tag\_orbit} and \texttt{try\_merge\_orbits}
preserve (\ref{etiquetas unicas}). This is clear for \texttt{try\_merge\_orbits},
since this function does not tag orbits with new types. So we focus
on the calls to \texttt{tag\_orbit}; let us begin with the one on
line \ref{tag_orbit achica} of Algorithm \ref{alg:open}. Notice
that immediately before this call is executed we check that the type
to be introduced does not occur anywhere in \texttt{orbits}, and thus
(\ref{etiquetas unicas}) is preserved. Finally, assume a call $\mathtt{tag\_orbit}(\bar{a},\itya a,\iuna a)$
on line \ref{tag_orbit no achica}. Due to the tests carried out before
this call we know that (at the time of this call) the top of the stack
is $(\sg(\bar{a}),L,G)$ and that no tuple in $G$ has the same type
as $\bar{a}$. We want to show that no orbit is tagged with $\itya a$.
For the sake of contradiction assume there is $\bar{b}$ such that
its orbit, say $O'$, has type $\itya a$. So, since no tuple from
$G$ is in $O'$, item (\ref{si no achica y no esta conectada todas etiquetadas})
says that every tuple from $\sg(\bar{b})$ has known type. It follows
that $\sg(\bar{b})\neq A$, since otherwise $\bar{a}$ would have
had known type. Hence $(\sg(\bar{a}),L,G)$ was pushed on the stack
by processing a tuple $\bar{g}$ which was found to have a new type.
But this is a contradiction, since $\sg(\bar{b})$ contains an isomorphic
copy of $\bar{g}$, and thus would have had known type.

\end{proof}

\begin{proposition}

Algorithm \ref{alg:open} is sound and complete.

\end{proposition}

\begin{proof}
Note that the algorithm terminates if and only if either:
\begin{itemize}
\item a call to\texttt{ try\_merge\_orbits} returns $\false$ and Algorithm
\ref{alg:open} returns $\false$, or
\item the stack is empty and Algorithm \ref{alg:open} returns $\true$.
\end{itemize}
In the former case, it is clear from an inspection of Algorithm \ref{alg:propagar},
that \texttt{try\_merge\_orbits} returns $\false$ if and only if there
are two orbits tagged with the same isomorphism type and different
relTypes. Thus, by Lemma \ref{lem:invariantes orbitas}, there are
tuples $\bar{a},\bar{b}$ such that $\bar{a}\approx\bar{b}$ and $\rty(\bar{a},R_{0},\ldots,R_{l-1})\neq\rty(\bar{b},R_{0},\ldots,R_{l-1})$.
So, Corollary \ref{cor:lema semantico sin repeticiones} says that
$R$ is not qf-definable in $\mathbf{A}$.

Suppose next that the algorithm ends due to the stack being empty.
Recall that the stack is initialized with $(A,L_{0},\emptyset)$,
so, by Lemma \ref{propiedades del stack}.(\ref{todas etiquetadas}),
every tuple in $\bigcup_{k\in\spec(R)}A^{(k)}$ has known type when
the algorithm halts. We prove that (3) of Corollary \ref{cor:lema semantico sin repeticiones}
holds. Fix $k\in\spec(R)$ and take $\bar{a},\bar{b}\in A^{(k)}$
such that $\bar{a}\approx\bar{b}$. Now Lemma \ref{lem:invariantes orbitas}
together with Lemma \ref{propiedades del stack}.(\ref{etiquetas unicas})
guarantee that $\bar{a}$ and $\bar{b}$ are in the same orbit, and
thus item (\ref{relType correcto}) of Lemma \ref{lem:invariantes orbitas}
says that $\rty(\bar{a},R_{0},\ldots,R_{l-1})=\rty(\bar{b},R_{0},\ldots,R_{l-1})$.

\end{proof}

\section{Computing QfDefAlg using a splitting strategy}\label{sec:posta}

\begin{algorithm}
\caption{}

\label{alg:posta}
\begin{algorithmic}[1]
\Function{\texttt{$\mathtt{SplittingAlgorithm}$}}{$\mathbf{A}$, \texttt{target}}

    \State{$\mathtt{formula}$ = $\bot$} \Comment{Initialization of candidate formula}
    \State{$\mathtt{full\_blocks}$ = []}
    \State{$k = \ar(\texttt{target})$}
    \State{\texttt{tuples} = \{\texttt{t} for \texttt{t} in $A^{(k)}$\}} \Comment{Tuples without repeated elements}
    \State{\texttt{terms\_to\_process} = [$x_i$ for i in $\{0,\dots,\ar(\mathtt{target})-1\}$]}
    \State{\texttt{witnesses} = []}
    \State{\texttt{new\_witnesses} = []}
    \State{\texttt{local\_formula} = $\top$}
    \State{\texttt{step} = 0} % se usa solo para la prueba
    
    \Comment{Now we create the structure with fields for the Block}
    \State{$B_0$ = Block$(\texttt{tuples}, \texttt{witnesses}, \texttt{new\_witnesses}, \texttt{terms\_to\_process},$}\label{line:creo_block}
    \Statex{\hspace{2.35cm}\texttt{local\_formula}, \texttt{step})}
    \State{\texttt{blocks\_to\_process} = [$B_0$]} \Comment{First block in the stack}
    \While{\texttt{blocks\_to\_process} $\neq$ []}
        \State{$B$ = pop(\texttt{blocks\_to\_process})}
        \If{$B$.$\mathtt{tuples}$ $\subseteq$ \texttt{target}} \label{line:iffull} \Comment{$B$ is a full block}
            \State{$\mathtt{formula}$ = $\mathtt{formula}$ $\lor$ $B$.$\mathtt{formula}$}
            \State{$\mathtt{full\_blocks}$.append($B$)}
            \State{\textbf{continue}} \label{line:fifull}
        \ElsIf{\texttt{$B$}.$\mathtt{tuples}$ $\cap$ \texttt{target} = $\emptyset$} \Comment{$B$ is a disposable block} \label{line:ifdisposable}
            \State{\textbf{continue}}  \label{line:fidisposable}
        \Else \Comment{$B$ is a mixed block}
            \If{$B$.\texttt{terms\_to\_process} == [] and $B$.\texttt{new\_witnesses} == []}
                \State{\Return{(\textbf{False}, $B$)}} \Comment{$B$ is a mixed terminal block, so $B$ is counterexample} \label{line:contraejemplo}
            \EndIf
            \State{\texttt{blocks\_to\_process} = process\_mixed\_block($B$) + \texttt{blocks\_to\_process}}
        \EndIf
    \EndWhile
    \State{\Return{(\textbf{True},$\mathtt{formula}$)}}\label{line:formula_patron}

\EndFunction

\end{algorithmic}
\end{algorithm}

We now present our second algorithm to decide quantifier-free definability, also based on Corollary \ref{cor:lema semantico sin repeticiones}. We call it the \emph{Splitting Algorithm}. It starts in the same way as the Merging Algorithm by preprocessing the target relation (see subsection \ref{preprocesamiento}), and breaking it down into several new targets of the form $\left\lfloor R_{\pat\bar{a}}\right\rfloor$ for some $\bar{a}$ in the original target relation (in particular its tuples have no repeated entries). The next step is to process these new targets grouped by arity. If any of these parts turns out to be undefinable the algorithm returns $\false$ (together with a counterexample). If on the other hand, all the parts, for all arities, are definable, then the algorithm returns $\true$, together with a defining formula.

For the sake of simplicity, the presentation that follows describes the algorithm processing a single part of the original target (instead of all the parts of a same given arity). The pseudocode for this is exhibited in Algorithm \ref{alg:posta}. The reader should have no trouble in understanding how to modify this algorithm to obtain one that works in the general case. Perhaps, the only caveat is how the final formula is produced, more on that below.

\subsection{Breakdown of the Splitting Algorithm}

In contrast to the merging strategy, where we completely computed the isomorphism type of every tuple first, and only then partitioned the set of tuples accordingly, this new algorithm starts with the whole set of $k$-tuples (with no repetitions), where $k$ is the arity of the target relation, forming a single block, and as the algorithm progresses this partition is refined. The key idea is that, at any given time, the tuples in a block have not yet found to be non-isomorphic. A step of the algorithm consists in evaluating a term on all tuples in a block, and comparing the resulting values with the values of previously evaluated terms. Then, the tuples in the block are partitioned into successor blocks, according to what value they evaluate to. In order to do this, a block carries the information of the terms already evaluated, and the information needed to generate the upcoming terms to be evaluated. To store these annotated blocks the algorithm employs a data type, called \texttt{Block}, which can be thought of as a dictionary with the following five fields:.

\begin{itemize}
    \item a set $\mathtt{tuples}$ containing the tuples of the block,
    \item three lists of terms $\mathtt{witnesses}$, $\mathtt{new\_witnesses}$, $\mathtt{terms\_to\_process}$,
    \item and $\mathtt{formula}$, containing a quantifier-free formula.
\end{itemize}

The lists of terms \texttt{witnesses} and \texttt{new\_witnesses} serve a purpose analogous to the variables \texttt{H} and \texttt{N}, respectively, in Algorithm \ref{alg:hit}.
The terms to be evaluated, in the current block and its successors, are generated in single chunk for each given depth $d$. To avoid generating redundant terms, we only build terms that include at least a term of depth $d-1$ that produced a new value in the last round. These are stored in $\mathtt{new\_witnesses}$, and avoiding the generation of redundant terms is the only purpose of this field. To understand the role of the field \texttt{witnesses} let us take a look at what happens when a block $B$ is processed. 
We say that a block is \emph{pure} provided it is either contained in or disjoint with the target. Since Algorithm \ref{alg:posta} is essentially trying to find a counterexample, that is, a pair of isomorphic tuples separated by the target relation, pure blocks need not be further inspected in search of said counterexample. Blocks that are disjoint with the target, called \emph{disposable}, do not need any further treatment. However, blocks contained in the target, which we call \emph{full}, contribute their \texttt{formula} as a disjunct of the output formula. Thus, Algorithm \ref{alg:posta} (lines \ref{line:iffull} to \ref{line:fidisposable}) begins the treatment of a block by checking whether it is full or disposable. Next, we analyze how blocks that are not pure, called \emph{mixed}, are processed. Let $B$ be a mixed block. We first consider the case where $B.\mathtt{terms\_to\_process}$ and $B.\mathtt{new\_witnesses}$ are empty. Such a block is called \emph{terminal}. This happens exactly (Lemma \ref{thm:hitcompleto}, (2)) when all tuples in the block are isomorphic, and, as $B$ is mixed, we have found a counterexample and the algorithm halts (line \ref{line:contraejemplo}). Assume next that either $B.\mathtt{terms\_to\_process}$ or $B.\mathtt{new\_witnesses}$ are nonempty. Here Algorithm \ref{alg:posta} calls the function \texttt{process\_mixed\_block} on $B$. On execution, this function first checks if there are terms to process in $B$. If that is not the case, then a new list of terms (of increased depth) is generated and assigned to $B.\mathtt{terms\_to\_process}$. On the other hand, if $B.\mathtt{terms\_to\_process}$ is not empty, we take its first element $t$. Next, the tuples in $B$ are distributed among successor blocks according to their value via $t$; for each $s\in B.\mathtt{witnesses}$ a new block is created with all tuples  $\bar{a}\in B$ such that $t^\mathbf{A}(\bar{a})=s^\mathbf{A}(\bar{a})$, assuming there is at least one such tuple. In addition, one further block is created collecting all tuples whose value via $t$ does not agree with any of the values produced by witnesses. It should be clear from the pseudocode how the rest of fields in the successor blocks are instantiated. The function returns the list of successor blocks of $B$ which in turn, is pushed onto the stack of blocks to process by the main algorithm. Eventually, we either encounter a terminal mixed block, or run out of blocks to process, case in which the algorithm halts returning the defining formula.

We conclude the breakdown of the Splitting Algorithm with a brief explanation of how the formula for the original target is constructed in the positive case. Recall that, on input $(\mathbf{A},R)$, we start by breaking down $R$ into relations of the form $\left\lfloor R_{\theta}\right\rfloor$ for some pattern $\theta$ present in $R$. In the case that $R$ is definable in $\mathbf{A}$, running the above algorithm produces a formula $\varphi_{\theta}$ for each $R_{\theta}$. Now, $\varphi_{\theta}$ employs the variables $x_0,\dots ,x_{|\theta|-1}$.  Note that $|\theta|$ is smaller than $k := \ar(R)$ unless $\theta = \{\{0\},\dots,\{k-1\}\}$, thus we need to substitute the variables in $\varphi_{\theta}$ for the corresponding variables from $\{x_0,\dots,x_{k-1}\}$, and add the (non)equalities between variables that describe the pattern $\theta$. This is best understood with an example. Suppose $k=5$ and $\theta = \{\{0,1\},\{2,3\},\{4\}\}$. In this case the resulting formula is 
\[
\tilde{\varphi}_{\theta}(x_0,x_1,x_2,x_3,x_4) :=  \varphi_{\theta}(x_0,x_2,x_4) \wedge x_0 = x_1 \wedge x_2 =x_3 \wedge x_0 \neq x_2 \wedge x_0 \neq x_4 \wedge x_2 \neq x_4.
\]
Now, if $\Theta$ is the set off all patterns present in $R$, we have that the quantifier-free formula defining $R$ in $\mathbf{A}$ is 
\[
\varphi(x_0,\dots ,x_{k-1}) := \bigvee_{\theta \in \Theta}\tilde{\varphi}_{\theta}(x_0,\dots ,x_{k-1}),
\]
where $\tilde{\varphi}_{\theta}$ is the modified formula for $\varphi_{\theta}$ for each $\theta \in \Theta$.

\begin{algorithm}

\caption{}
\label{alg:procmix}
\begin{algorithmic}[1]

\Function{\texttt{$\mathtt{process\_mixed\_block}$}}{$B$}
    \State{$\mathtt{successors}$ = []}
    %\State{B.actualizar\_HIT()}
    \If{B.$\mathtt{terms\_to\_process}$ == []}  \Comment{Need to compute a new group of terms} \label{ifttpvacio} 
        \State{Generate terms using B.$\mathtt{witnesses}$ and B.$\mathtt{new\_witnesses}$ in B.$\mathtt{terms\_to\_process}$ } \label{nuevosaprocesar}
        \State{B.$\mathtt{new\_witnesses}$ = []} \label{fittpvacio}
    \EndIf
    \State{$\mathtt{t}$ = pop(B.$\mathtt{terms\_to\_process}$)}
    \State{B.$\mathtt{step}$ = B.$\mathtt{step}$ + 1} \label{line:step+1}
    \State{$\mathtt{complement\_block}$ = B.$\mathtt{tuples}$} \Comment{Initialize with all tuples}
    \State{$\mathtt{complement\_formula}$ = $\top$}
    \For{s $\in$ B.$\mathtt{witnesses}$} \label{line:forsplit}
        \State{$\mathtt{eq\_tuples}$ = [v $\in$ B.$\mathtt{tuples}$ : $\mathtt{t}$(v) == s(v)]} \Comment{Tuples for a new block}
        \If{$\mathtt{eq\_tuples}$ != []}
            \State{$\mathtt{successor}$ := Block($\mathtt{eq\_tuples}$, B.$\mathtt{witnesses}$, B.$\mathtt{new\_witnesses}$,}
            \Statex{\hspace{4.7cm} B.$\mathtt{terms\_to\_process}$, B.$\mathtt{formula}$ $\land$ $t=s$, B.$\mathtt{step}$)}
            \State{$\mathtt{successors}$.append($\mathtt{successor}$)}
            
            \Comment{Delete the tuples that appear in the new block}
            
            \State{$\mathtt{complement\_block}$ = $\mathtt{complement\_block}$ - $\mathtt{eq\_tuples}$} \label{line:deletetuples}

            \Comment{Append the negation formula}
            
            \State{$\mathtt{complement\_formula}$ = B.$\mathtt{formula}$ $\land$ $\mathtt{complement\_formula}$ $\land$ $(t\neq s)$ }
        \EndIf
    \EndFor
    \If{$\mathtt{complement\_block}$ != []}
        \State{$\mathtt{successor}$ = Block($\mathtt{complement\_block}$, B.$\mathtt{witnesses}$ + [$\mathtt{t}$],}
         
         \Statex{\hspace{4.1cm}B.$\mathtt{new\_witnesses}$ + [$\mathtt{t}$], B.$\mathtt{terms\_to\_process}$,}
        \Statex{\hspace{4.051cm} $\mathtt{complement\_formula}$, B.$\mathtt{step}$)}
        \State{$\mathtt{successors}$.append($\mathtt{successor}$)} \label{line:endsplit}
    \EndIf
    \If{there is exactly one entry in $\mathtt{successors}$}
        \Comment{$\mathtt{formula}$ does not need to increase} \label{line:unhijo}
        \State{$\mathtt{successors}$[0].$\mathtt{formula}$ = B.$\mathtt{formula}$}
    \EndIf
    \State{\Return{$\mathtt{successors}$}}\label{line:returnsuccessors}
\EndFunction

\end{algorithmic}
\end{algorithm}

\bigskip
\bigskip

The variables $\mathtt{step}$ and $\mathtt{full\_blocks}$ are not necessary for the algorithm itself, but are included to make the proofs easier to follow.

\subsection{Soundness and completeness}
In this section we aim to prove correctness of the Splitting Algorithm. We start with some preliminary properties.

Our proof is based on the following consequence of Theorem \ref{thm:lema sem}. Recall that given $\bar{a},\bar{b} \in A^k$, we say $\bar{a} \approx \bar{b}$ when $t^{\mathbf{A}}(\bar{a})=s^{\mathbf{A}}(\bar{a})\Longleftrightarrow t^{\mathbf{A}}(\bar{b})=s^{\mathbf{A}}(\bar{b})$ for every pair of terms $t,s$.

\begin{corollary}

\label{corolario semantico}Let $\mathbf{A}$ be a finite algebra and $R$
a $k$-ary relation over $\mathbf{A}$. Then the following are equivalent:
\begin{enumerate}
\item $R$ is not definable in $\mathbf{A}$.
\item There are tuples $\bar{a},\bar{b} \in \mathbf{A}^{k}$ such that $\bar{a} \in R,\,\bar{b} \notin R$
and $\bar{a} \approx \bar{b}$ 
\end{enumerate}
\end{corollary}

Given a term $t(x_{0},\ldots,x_{k-1})$, define $\depth(t)$ to be
the least natural number $i$ such that $t\in\mathcal{T}_{i}(x_{0},\ldots,x_{k-1})$. 
Also, for a block $B$, we define $\depth(B)=\max\{\depth(t):t\in B.\mathtt{witnesses} \dobleplus B.\mathtt{terms\_to\_process}\}$, where $\dobleplus$ is the list concatenation operator. Recall that, given a formula $\varphi$ and a model $\mathbf{A}$, the extension of $\varphi$ in $\mathbf{A}$ is denoted by $[\varphi]^\mathbf{A}$.  

\begin{lemma}

\label{thm:invariantes}Let $\mathbf{A}$ be a finite algebra and $R$
a $k$-ary relation over $\mathbf{A}$. Then the following are invariant throughout the execution
of Algorithm \ref{alg:posta} on input $(\mathbf{A},R)$:

\begin{enumerate}
\item For any block $B$, $t,s\in B.\mathtt{witnesses}$ and $\bar{a}\in B.\mathtt{tuples}$, we have $t(\bar{a})=s(\bar{a})$ implies $t=s$. \label{termsnorepiten}
\item The sets of tuples of the blocks in $\mathtt{blocks\_to\_process} \dobleplus \mathtt{full\_blocks}$ are pairwise disjoint and their union contains $R$. \label{bloquesdisjuntos}
\item For any block $B$, we have $[B.\mathtt{formula}]^\mathbf{A}=B.\mathtt{tuples}$ \label{extensiondeB.formula}
\item For any block $B$ and for every term  $t$ in
$\mathcal{T}_{d}(x_{0},\ldots,x_{k-1})$, where  $d=\depth(B)$, there is a term $\hat{t}$
in $B.\mathtt{witnesses} \dobleplus B.\mathtt{terms\_to\_process}$ such that $t(\bar{a})=\hat{t}(\bar{a})$ for every $\bar{a}$
in $B.\mathtt{tuples}$. \label{terminosrepresentados}
\item For any block $B$ and for every term  $t$ in
$\mathcal{T}_{d}(x_{0},\ldots,x_{k-1}) \setminus \mathcal{T}_{d-1}(x_{0},\ldots,x_{k-1})$, where  $d=\depth(B)$, there is a term $\hat{t}$
in $B.\mathtt{new\_witnesses} \dobleplus B.\mathtt{terms\_to\_process}$ such that $t(\bar{a})=\hat{t}(\bar{a})$ for every $\bar{a}$
in $B.\mathtt{tuples}$. \label{terminosdecaparepresentados}
\item For any $\bar{a},\bar{b}\in A^{(k)}$
such that $\bar{a}\approx\bar{b}$, there is a block $B$ for which
$\bar{a},\bar{b}\in B.\mathtt{tuples}$. \label{tuplasisomorfasjuntas}
\end{enumerate}
\end{lemma}

\begin{proof}

(\ref{termsnorepiten}) This follows from simply observing on the pseudocode that a popped
term $t$ is appended to a list of terms in a block $B$ being constructed only if $t(\bar{a})\neq s(\bar{a})$
for all $s\in B.\mathtt{witnesses}$, $\bar{a}\in B.\mathtt{tuples}$.

(\ref{bloquesdisjuntos}) Observe that \texttt{process\_mixed\_block} does not split the block being processed
when the block is moved from $\mathtt{blocks\_to\_process}$ to $\mathtt{full\_blocks}$ (Algorithm \ref{alg:posta}, lines \ref{line:iffull} to \ref{line:fifull})
or when $\depth(B)$ is increased (Algorithm \ref{alg:procmix}, lines \ref{ifttpvacio} to \ref{fittpvacio}), so we just need to prove the invariant remains true when $\mathtt{step}$ is increased (Algorithm \ref{alg:procmix}, lines \ref{line:step+1} to \ref{line:returnsuccessors}). Since only disposable blocks are discarded, $R$ is always
contained in the union of blocks in $\mathtt{blocks\_to\_process} \dobleplus \mathtt{full\_blocks}$; from line \ref{line:deletetuples} in Algorithm \ref{alg:procmix}, it is straightforward these blocks remain pairwise disjoint.

(\ref{extensiondeB.formula}) The statement clearly holds for the initial block and in the following instances:
\begin{itemize}
\item when a full block is moved from \texttt{blocks\_to\_process} to \texttt{full\_blocks}
(Algorithm \ref{alg:posta}, lines \ref{line:iffull} to \ref{line:fifull})
\item when a disposable block is removed from \texttt{blocks\_to\_process} (Algorithm \ref{alg:posta}, lines \ref{line:fidisposable} to \ref{line:fidisposable})
\item when $\depth(B)$ is increased (Algorithm \ref{alg:procmix}, line \ref{nuevosaprocesar}) 
\item after a term $t$ is popped from \texttt{terms\_to\_process}, increasing \texttt{step}
but leaving $B$ unpartitioned (if the condition in Algorithm \ref{alg:procmix}, line \ref{line:unhijo} holds)
\end{itemize}
Let us now consider the case where the block $B$ is partitioned after
popping the term $t$ (Algorithm \ref{alg:procmix}, lines \ref{line:forsplit} to \ref{line:endsplit}). Assume inductively
that $\left[B.\mathtt{formula}\right]^{A}=B.\mathtt{tuples}$. By construction, on each
of the new blocks the statement becomes $\left[(B.\mathtt{formula})\wedge(t=s)\right]^{A}=\left[t=s\right]^{A}\cap B.\mathtt{tuples}$
for some $s$ in $B.\mathtt{witnesses}$, which follows immediately from the inductive
hypothesis. 

(\ref{terminosrepresentados}) and (\ref{terminosdecaparepresentados}) We proceed by induction on $d$. In the case $d=0$ and
$\mathtt{step}=0$ we have $B.\mathtt{witnesses}=[\ ]$ and $B.\mathtt{terms\_to\_process}=[x_{0},\ldots,x_{k-1}]$.
Notice we can assume there are no constant symbols in the language,
since they can be substituted by constant unary functions, which means
the statement is true in this case.

Assume now $d>0$ and $\mathtt{step}>0$. We note none of the instructions
after increasing $\mathtt{step}$ break the invariants: the blocks that arise
from the process following the popping of a term from $B.\mathtt{terms\_to\_process}$
inherit the content of $B.\mathtt{witnesses}$ and $B.\mathtt{terms\_to\_process}$, with
the exception perhaps of the popped term $t$ if it agrees with
some other term on all the tuples of the block to be created.

So we can actually assume $\mathtt{step}=0$. At this point, the algorithm
reached the current state after constructing a sequence of terms in
$B.\mathtt{terms\_to\_process}$ from a previous mixed block where
$\mathtt{terms\_to\_process}$ was emptied, increasing $d$ by $1$. Let $t=f(t_{0},...,t_{r-1})\in\mathcal{T}_{d+1}$.
By the inductive hypothesis, there are terms $\hat{t}_{0},...,\hat{t}_{r-1}\in\mathcal{T}_{d}$,
each satisfying $t_{i}^{\mathbf{A}}(\bar{a})=\hat{t_{i}}^{\mathbf{A}}(\bar{a})$,
and we can assume there is $i_{0}$ such that $t_{i_{0}} \in \mathcal{T}_{d}(x_{0},\ldots,x_{k-1}) \setminus \mathcal{T}_{d-1}(x_{0},\ldots,x_{k-1})$.
But this means $\hat{t}=f(\hat{t}_{0},...,\hat{t}_{r-1})\in B.\mathtt{terms\_to\_process}$.

(\ref{tuplasisomorfasjuntas}) As the algorithm starts, every such pair of tuples is in the original
block containing all tuples; when a block is partitioned, clearly
lines \ref{line:forsplit} to \ref{line:endsplit} place
the pair on the same block.

\end{proof}

\begin{lemma} \label{thm:hitcompleto}

Let $B$ be a block with $B.\mathtt{new\_witnesses}=B.\mathtt{terms\_to\_process}=[\,]$. Then:
\begin{enumerate}
    \item For any term $t$ there is a term $\hat{t}\in B.\mathtt{witnesses}$ such that
    $t(\bar{a})=\hat{t}(\bar{a})$ for every $\bar{a}\in B.\mathtt{tuples}$.
    \item If $\bar{a},\bar{b}\in B.\mathtt{tuples}$ and $t,s$ are any terms, then
    $t(\bar{a})=s(\bar{a})$ implies $t(\bar{b})=s(\bar{b})$.
\end{enumerate}
\end{lemma}

\begin{proof}
(1) Define $d=\depth(B)$ and let $t\in\mathcal{T}_{d}$ be a term.
If $\hat{t}$ is the term given by (\ref{terminosrepresentados}), then (\ref{terminosdecaparepresentados})
along with the hypotheses says $\hat{t}\in\mathcal{T}_{d-1}$. The
result now follows from this observation by an easy inductive argument.

(2) Take $\bar{a},\bar{b}\in B.\mathtt{tuples}$. For any terms $t,s$, let
$\hat{t},\hat{s}$ be the terms given by (1). Then $\hat{t}(\bar{a})=\hat{s}(\bar{a})$,
which means by Lemma \ref{thm:invariantes} (\ref{termsnorepiten}) that $\hat{t}=\hat{s}$, so $t(\bar{b})=\hat{t}(\bar{b})=\hat{s}(\bar{b})=s(\bar{b})$.

\end{proof}

\begin{proposition}

\label{correccion_posta} Algorithm \ref{alg:posta} is sound and complete. More
precisely, on input $(\mathbf{A},R)$, Algorithm \ref{alg:posta} eventually stops and it either returns a formula $\varphi$ satisfying $R=[\varphi]^\mathbf{A}$, or it returns a counterexample. A counterexample is returned iff $R$ is not qf-definable.

\end{proposition}

\begin{proof}

Let us first consider the case where $R$ is qf-definable. This means no mixed block $B$  will ever be terminal, which implies that every mixed block will eventually be split into (pure) full or disposable blocks. The fact that these are removed from
$\mathtt{blocks\_to\_process}$ and $A$ being finite warranty termination. Observe
also that $\mathtt{full\_blocks}$ contain only pure full blocks; therefore, as the algorithm stops by emptying $\mathtt{blocks\_to\_process}$, the union
set in (\ref{bloquesdisjuntos}) of Lemma (\ref{thm:invariantes}) coincides exactly with $R$ and the algorithm
returns a formula $\varphi$, which by construction is written in
normal disjunctive form. Each of its disjuncts corresponds
to a pure full block added to $\mathtt{full\_blocks}$, and by (\ref{extensiondeB.formula}) in Lemma (\ref{thm:invariantes}),
the tuples on each of these blocks form the extension of the corresponding
disjunct. By the above considerations, we have $R=[\varphi]^\mathbf{A}$.

If the algorithm did not return a formula $\varphi$, it means termination
took place because a counterexample was found; this happens exactly
when $R$ is not qf-definable by Corollary \ref{corolario semantico}.

\end{proof}

\section{Empirical comparisons of merging versus splitting algorithms\label{sec:testing}}

\newcommandx\relationalmodel[1][usedefault, addprefix=\global, 1=t]{\boldsymbol{\left\langle #1\right\rangle ^{Rel}}}%

Having two different algorithmic approaches to solve the quantifier-free definability problem, it is natural to wonder if either one outperforms the other. Such a comparison at a theoretical level turns out to be extremely difficult. Thus, our compromise solution was to run (implementations of) these algorithms on a series of test scenarios. As a byproduct of these tests we obtain a better understanding as regards the practical limitations of our implementations in terms of the input size.

Since there is no standard set of tests for the quantifier-free definability problem, we ran our algorithms on three types of algebraic structures: boolean
algebras, abelian groups and randomly generated algebras. These latter were generated by choosing random operations over a fixed universe. For each family we ran the test on 300 algebras and we took the median of the wall time samples.

Inspecting both algorithmic strategies it is not hard to conclude that the worst-case scenario is when the target relation is definable. This was further confirmed during testing, where we found that both algorithms take very little time to decide that a given target relation is not quantifier-free definable. Hence, all the tests presented are for inputs where the target is definable. 

Our implementations are written in Python 3 under GPL 3 license, source code for merging algorithm is available at \href{https://github.com/pablogventura/QfDefAlg}{https://github.com/pablogventura/QfDefAlg}, and for splitting algorithm at \href{https://github.com/pablogventura/QfDefAlgSplitting}{https://github.com/pablogventura/QfDefAlgSplitting}.
All tests were performed using an Intel Xeon E5-2620v3 processor with
12 cores (however, our algorithms do not make use of parallelization)
at 2.40GHz, 128 GiB DDR4 RAM 2133MHz. Memory was never an issue.

\subsection{Assessing the performance of the Merging Algorithm}

Our first set of empirical data concerns the performance of Algorithm \ref{alg:open} on various kinds of inputs. The
cardinalities of the algebras run through 4, 8, 16 and 32. In all cases
the target was the binary relation consisting of all pairs
from the domain of the algebra. Random algebras are endowed with a
binary and a ternary operation. Abelian groups are obtained as products
of cyclic groups $\mathbb{Z}_{k}$. 

\begin{figure}[h]
\caption{Time to decide definability using merging algorithm}\label{graph:megahit}
\centering

\includegraphics[width=0.6\textwidth]{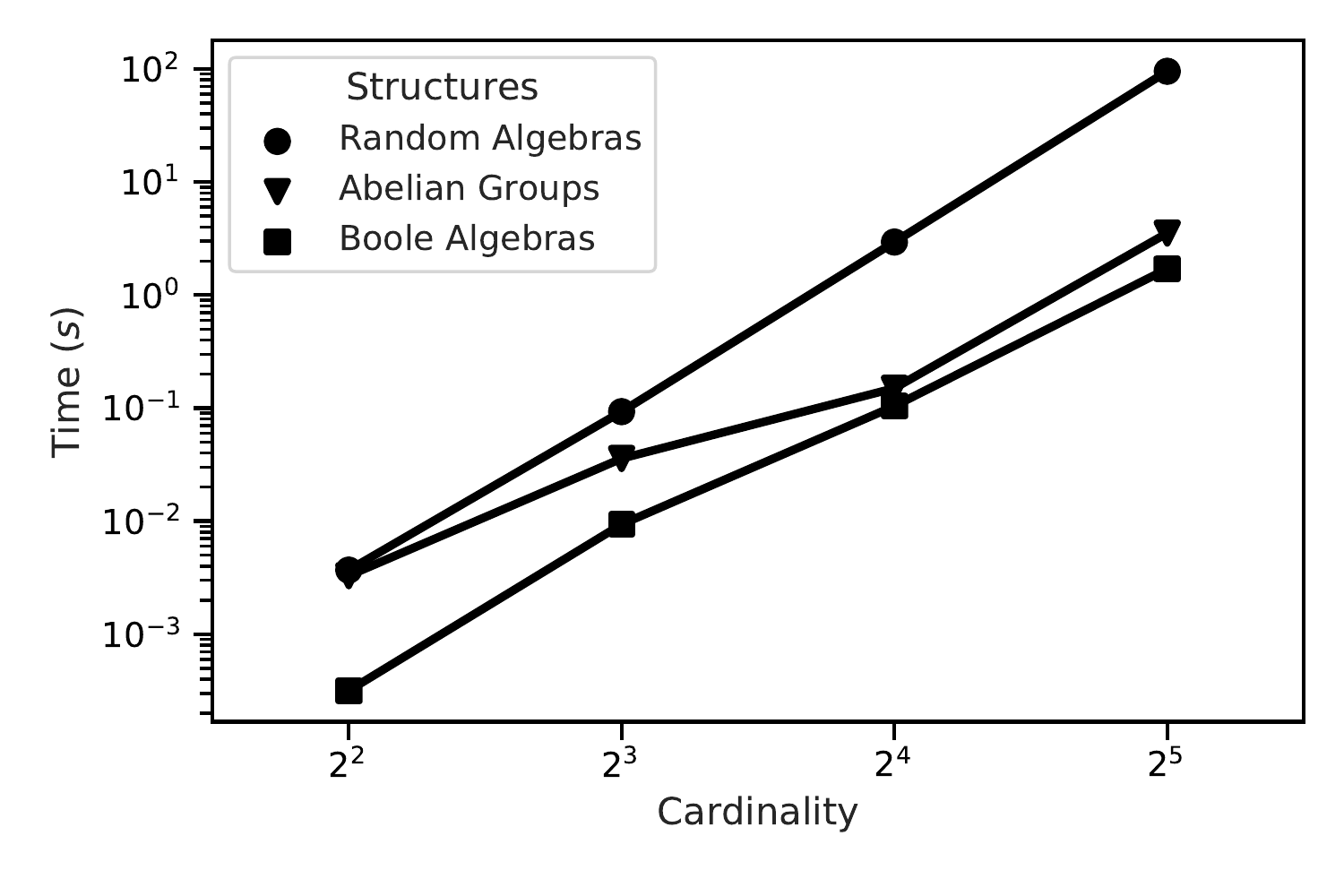}
\end{figure}

The results are exhibited in Figure \ref{graph:megahit}. We can see how the algorithm takes advantage of inner symmetries of
the algebras: the running time increases as the number
of subisomorphisms decreases. In particular, it is interesting to note the behaviour of the abelian groups curve. Since all groups in the test are
direct products with factors in $\{\mathbb{Z}_{2}, \mathbb{Z}_{4}\}$ (e.g., the 16-element one is $\mathbb{Z}_{2}\times\mathbb{Z}_{2}\times\mathbb{Z}_{4}$ and the 32-element one is  $\mathbb{Z}_{2}\times\mathbb{Z}_{4}\times\mathbb{Z}_{4}$),
the fact that the larger groups have repeated factors increases the number of subisomorphisms, and thus our algorithm has a better performance.

\subsection{Assessing the performance of the Splitting Algorithm}

\begin{figure}[h]
\caption{Time to decide definability using splitting algorithm}\label{graph:posta}
\centering

\includegraphics[width=0.6\textwidth]{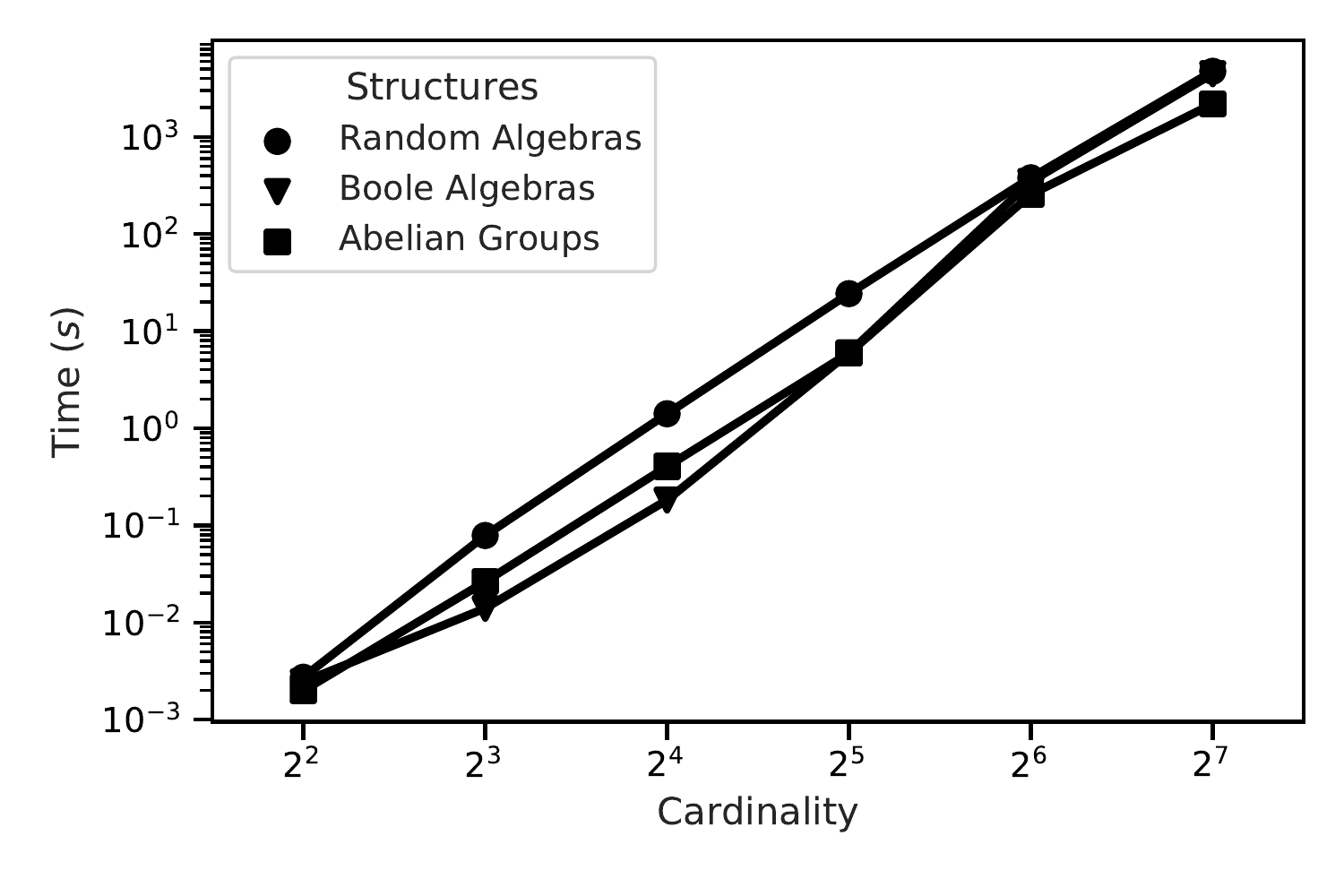}
\end{figure}

We now turn to the performance analysis of Algorithm \ref{alg:posta}. It turns out that this algorithm was able to handle input algebras of up to 128 elements. Even more striking is the fact that the target relations in these tests are ternary, given that the search space depends exponentially on the arity of the target (for more on this see \cite{complejidad}, where the parameterized complexity of $\mathrm{QfDefAlg}$ is studied). It is also worth mentioning that, since splitting algorithm produces an immediate answer on a target relation containing all tuples of a given length, the input relation for these tests was obtained by taking the extension of randomly constructed formula.

In Figure \ref{graph:posta}, we see that, even when structures with more inner symmetries continue to take less to be processed, the difference is now not as noticeable. This can be explained by the fact that the Splitting algorithm does not need to completely determine the isomorphism types of every tuple in the search space.

\subsection{Comparing both approaches}

\begin{figure}[h]
\caption{Times using merging vs splitting algorithm}\label{graph:postavsmegahit}
\centering

\includegraphics[width=0.6\textwidth]{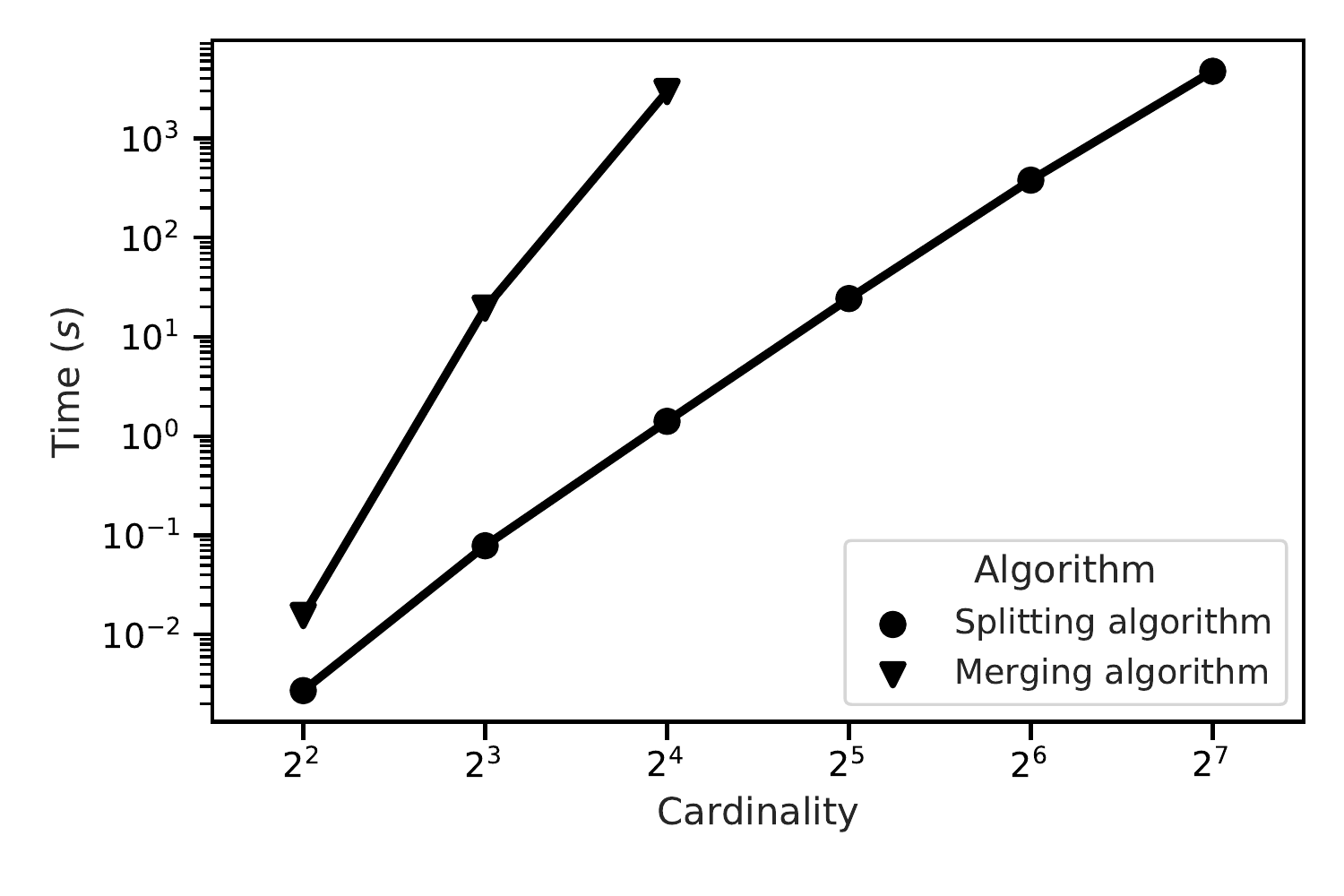}
\end{figure}

Our final set of empirical data shows the performance of both algorithms on the same sample of inputs. Here, as we did in the tests for the Splitting algorithm, the input relation is obtained as the extension of a randomly generated quantifier-free formula. The results are presented in Figure \ref{graph:postavsmegahit}. As anticipated by our previous tests, the Splitting algorithm greatly outperforms the Merging algorithm. In fact, we can see the Merging algorithm falls out of the considered time-window on models of size 32 and up (projections foretell execution times over 25 hours). An obvious reason that explains this striking difference in performance is provided by the fact that, in contrast to the Merging algorithm, the Splitting algorithm does not compute the isomorphism type of every tuple in the search space. In addition, for many of the tuples processed by the Splitting algorithm the isomorphism type is only partially computed.
Considering the Splitting algorithm also provides a formula in the positive case, we can surmise that the Splitting algorithm is clearly a more efficient strategy.

\section{Conclusions and future work\label{sec:conclusiones}}

We have presented two algorithms that decide the definability problem
by quantifier-free first-order formulas over a purely functional language. These
algorithms rely on the semantic characterization of quantifier-free definability
given in \cite{camp:lemas_semanticos}. We showed that this problem
is coNP-complete, which is the same complexity we obtained in the relational case \cite{complejidad}.
Prior to this we developed an algorithm that gives a characterization
of the isomorphism type of a tuple $\bar{a}$ in the given structure
(in particular, the subuniverse it generates is obtained).

Our first definability-decision procedure (Merging Algorithm) partitions the set of all $k$-tuples
from the universe with $k\in\spec(R)$, and has to exhaust the search
space to give a positive answer. Therefore, the execution time of
our algorithm depends exponentially on the parameter $k$. The empirical tests confirm the exponential dependence on $k$, and give us an idea of the impact of the cardinality of the input structure and the family of algebras on which we performed the tests. They also confirm the conjecture that this strategy, which is based on the existence of symmetries, would perform better on models with a large amount of subisomorphisms, as we 

The Splitting algorithm starts with a single block of tuples, and partitions blocks as the determination of the isomorphism types of their tuples progresses. This differs from the previous strategy, where we started with isolated tuples and merged those blocks with the same isomorphism type. As one could expect, this new strategy proved much more efficient, since entire blocks can be discarded without completely determining the isomorphism type of their tuples. The superior performance was clearly confirmed by the empirical testing. In addition to this, the fact that this approach allows the defining formula to be obtained in the positive case constitutes a fundamental advantage.

There are several possible lines for future research. Due to the lack
of standard testings to decide definability, the empirical tests we
presented are preliminary, and more evaluation is needed. The formula produced by the Splitting algorithm (in the positive case) turned out to be very large in testing. Thus, an interesting problem that arises is to understand how to simplify (if possible) the produced formula. This appears to be a difficult problem, and we know that in some cases defining formulas are necessarily large (see \cite{complejidad}).
A seemingly straightforward modification we have planned is to extend the algorithms in this paper to decide definability over finite classes of models instead of a single structure. Lastly, it would be interesting to identify classes of algebras in which the problem is well-conditioned and for which polynomial time algorithms exist.

\bibliographystyle{plain}
\bibliography{references}

\end{document}